# Hand-waving and Interpretive Dance:
# An Introductory Course on Tensor Networks
# **Lecture Notes**

Jacob C. Bridgeman[1], Christopher T. Chubb[2]

Centre for Engineered Quantum Systems
School of Physics, The University of Sydney, Sydney, Australia

Australian Institute of Nanoscale Science and Technology
Sydney Nanoscience Hub, The University of Sydney, Sydney, Australia

Version 4, May 17, 2017

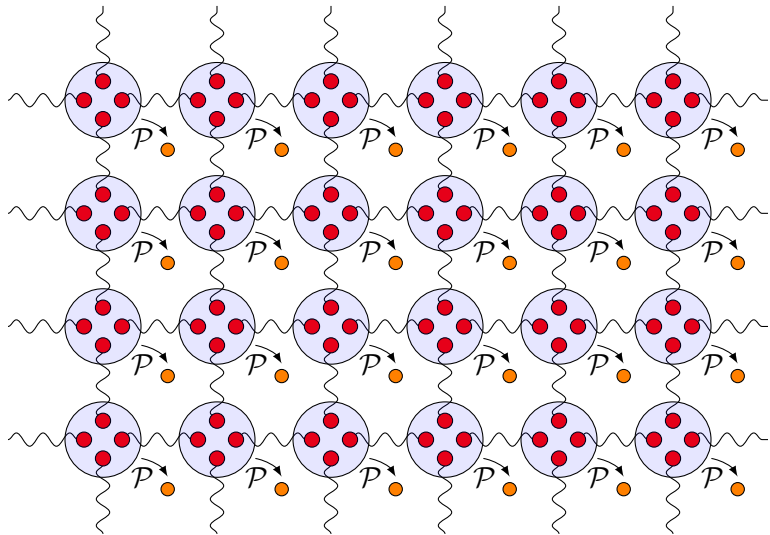

[1] jacob.bridgeman@sydney.edu.au, http://www.physics.usyd.edu.au/~jbridge
[2] christopher.chubb@sydney.edu.au, http://www.physics.usyd.edu.au/~cchubb

## Abstract

The curse of dimensionality associated with the Hilbert space of spin systems provides a significant obstruction to the study of condensed matter systems. Tensor networks have proven an important tool in attempting to overcome this difficulty in both the numerical and analytic regimes.

These notes form the basis for a seven lecture course, introducing the basics of a range of common tensor networks and algorithms. In particular, we cover: introductory tensor network notation, applications to quantum information, basic properties of matrix product states, a classification of quantum phases using tensor networks, algorithms for finding matrix product states, basic properties of projected entangled pair states, and multiscale entanglement renormalisation ansatz states.

The lectures are intended to be generally accessible, although the relevance of many of the examples may be lost on students without a background in many-body physics/quantum information. For each lecture, several problems are given, with worked solutions in an ancillary file.

## Acknowledgements

We thank everyone who stayed awake for the presentation of these lectures in January of 2016 at the Australian Institute of Nanoscience. We thank Stephen Bartlett, Andrew Doherty, Christopher Granade, Robin Harper, Marco Tomamichel, Dominic Williamson, and especially Doriane Drolet and David Tuckett, for their input. For suggesting that we give these lectures and editorial assistance, we give special thanks for the best-selling childrens' author[1] Chris Ferrie. We acknowledge support from the Australian Research Council via the Centre of Excellence in Engineered Quantum Systems (EQuS), project number CE110001013.

Much of the material was reproduced from memory after one of the authors attended the Tensor Network Summer School at Universiteit Gent in 2015.

---

# Contents

# 0    Introduction

One of the biggest obstacles to the theoretical and numerical study of quantum many-body systems is the *curse of dimensionality*, the exponential growth of the Hilbert space of quantum states. In general this curse prevents efficient description of states, providing a significant complexity barrier to their study. Despite this, physically relevant states often possess additional structure not found in arbitrary states, and as such *do not* exhibit this pathological complexity, allowing them to be efficiently described and studied.

Tensor networks have proven to be an incredibly important technique in studying condensed matter systems, with much of the modern theory and numerics used to study these systems involving tensor networks.

In the numerical regime, tensor networks provide variational classes of states which can be efficiently described. By, for example, minimising the energy over one of these classes, one can learn a great deal about the low-energy behaviour some physical system of interest. The key variational classes are: matrix product states (MPS), projected entangled pair states (PEPS), and multiscale entanglement renormalisation ansatz (MERA). Due to their importance, and prevalence in the literature, we devote a chapter to each of these.

By studying the structure and properties of classes tensor networks, for example MPS, one can learn a great deal about the types of states which they can describe. Tensor network states therefore provide an important analytic framework for understanding the universal properties of classes of states which possess particular properties, such as those which only support certain entanglement or correlation structures.

In addition to their application to many-body physics, tensor networks can also be used to understand many of the foundational results in quantum information. The understanding of concepts such as quantum teleportation, purification, and the church of the larger Hilbert space, can be understood relatively simply when the tensor network framework is utilised. Some examples of this are presented in Section 2. These lectures aim to introduce, and make familiar, the notation conventionally used for tensor network calculations. As a warm up, we present some key quantum information results in this notation.

After introducing the class of MPS, we present some of the key properties, as well as several analytic matrix product states examples, which can serve as useful toy models. To demonstrate the analytic power of MPS we will then consider a key result in condensed matter theory: the classification of one-dimensional phases. This serves as an example of a result which, within the tensor network formalism, can be much more succinctly and clearly explained than it can in more standard linear algebraic notation.

When utilising tensor networks numerically, algorithms must be designed which, for example, minimise the energy of some Hamiltonian over the variational class. We introduce two such algorithms, namely DMRG and TEBD, which are particularly prevalent. These have become standard tools in numerical many-body physics.

We then introduce the class of PEPS, a class designed for two-dimensional many-body systems. We discuss some of the properties, and some of the challenges to simulating using this class of networks.

Finally, we introduce another class, MERA, which can be utilised for the study of gapless one-dimensional (and higher!) systems. This class has many interesting properties, including an interpretation as a renormalisation group. This has sparked interest in a wide range of field, from quantum information to string theory.

# 1 Introduction to Tensor Network Notation

One of the primary reasons that tensor networks are so useful is the straightforward and transparent notation usually used to describe them. Using a graphical language, the structure is manifest. Many general properties of the objects under study, particularly quantum states, can be identified directly from the structure of the network needed to describe them.

Tensor network notation (TNN) can be considered a generalisation of Einstein summation notation. In this lecture we will define tensor networks, starting with an introduction to tensors and the operations we can perform upon them.

## 1.1 Tensors

Tensors are a generalisation of vectors and matrices. A $d$-dimensional vector can be considered an element of $\mathbb{C}^d$, and a $n \times m$-dimensional matrix an element of $\mathbb{C}^{n \times m}$. Correspondingly a rank-$r$ tensor of dimensions $d_1 \times \cdots \times d_r$ is an element of $\mathbb{C}^{d_1 \times \cdots \times d_r}$. We can clearly see that scalars, vectors and matrices are all therefore rank 0, 1 and 2 tensors respectively.

In tensor network notation a single tensor is simply represented by a geometric shape with legs sticking out of it, each corresponding to an index, analogous to the indices of Einstein notation. For example a rank-four tensor $R$ would be represented as

$$R^{\rho}_{\ \sigma\mu\nu} \implies \raisebox{-0.5em}{\textcircled{$R$}} \ . \tag{1.1}$$

In some contexts the shape used and direction of the legs can imply certain properties of the tensor or index — for a general network however, neither carry any special significance. When representing quantum states, it is often convenient to use the direction of legs to denote whether the corresponding vectors live in the Hilbert space ('kets') or its dual ('bras'). By adhering to this convention, certain prohibited contractions can be easily disallowed, such as contraction between two kets. This is notationally analogous to the convention of upper and lower denoting co- and contra-variant indices in Einstein or Penrose notation (a specialised form of TNN) employed in the study of general relativity or quantum field theory.

Because quantum mechanics, in contrast to general relativity, is complex, care has to be taken with complex conjugation. This is usually indicated either by explicitly labelling the tensor or adopting some index convention, such as flipping a network (upward and downward legs being echanged) carrying an implicit conjugation.

## 1.2 Tensor operations

The main advantage in TNN comes in representing tensors that are themselves composed of several other tensors. The two main operations we will consider are those of the tensor product and trace, typically used in the joint operation of contraction. As well as these two operations, the rank of a tensor can be altered by grouping/splitting indices.

### 1.2.1 Tensor product

The first operation we will consider is the tensor product, a generalisation of the outer product of vectors. The value of the tensor product on a given set of indices is the element-wise product of the values of each constituent tensor. Explicitly written out in index notation, the binary tensor product has the form:

$$[A \otimes B]_{i_1,\ldots,i_r,j_1,\ldots,j_s} := A_{i_1,\ldots,i_r} \cdot B_{j_1,\ldots,j_s} . \tag{1.2}$$

Diagrammatically the tensor product is simply represented by two tensors being placed next to each other. As such the value of a network containing disjoint tensors is simply the product of the constituent values.

$$A \otimes B \quad (1.3)$$

### 1.2.2 Trace

The next operation is that of the (partial) trace. Given a tensor $A$, for which the $x$th and $y$th indices have identical dimensions ($d_x = d_y$), the partial trace over these two dimensions is simply a joint summation over that index:

$$[\text{Tr}_{x,y}\, A]_{i_1,\ldots,i_{x-1},i_{x+1},\ldots,i_{y-1},i_{y+1},\ldots,i_r} = \sum_{\alpha=1}^{d_x} A_{i_1,\ldots,i_{x-1},\alpha,i_{x+1},\ldots,i_{y-1},\alpha,i_{y+1},\ldots,i_r} \quad (1.4)$$

Similar to Einstein notation, this summation is implicit in TNN, indicated by the corresponding legs being joined. An advantage over Einstein notation is that these summed-over indices need not be named, making the notation less clunky for large networks. For example, consider tracing over the two indices of a rank-3 tensor:

$$\text{---}\bigcirc\hspace{-0.3em}\supset \;\; := \text{Tr}_{\text{Right}}\left(\text{---}\bigcirc\hspace{-0.5em}\big<\right) = \sum_i \;\text{---}\bigcirc\hspace{-0.3em}{}^{\,i}_{\,i} \quad (1.5)$$

One property of the trace we can trivially see from this notation is that of its cyclic property. By simply sliding one of the matrices around – which only changes the placement of the tensors in the network, and therefore not the value – we can cycle the matrices around (being careful of transpositions), proving $\text{Tr}(AB) = \text{Tr}(BA)$.

$$\text{Tr}(AB) = \;\overbrace{\boxed{A}\text{---}\boxed{B}}\; = \;\overbrace{\boxed{\overset{\text{\rotatebox{180}{$B$}}}{\boxed{A}}}}\; = \;\overbrace{\boxed{\overset{B^T}{\boxed{A}}}}\; = \;\overbrace{\boxed{B}\text{---}\boxed{A}}\; = \text{Tr}(BA) \quad (1.6)$$

Whilst this serves as a trivial example, the higher rank equivalents of this statement are not always so obvious, and the fact that these properties hold 'more obviously' in TNN is often useful.

### 1.2.3 Contraction

The most common tensor operation used is *contraction*, corresponding to a tensor product followed by a trace between indices of the two tensors. An example would be the contraction between two pairs of indices of two rank-3 tensors, which is drawn as:

$$\text{---}\bigcirc\hspace{-0.3em}\bigcirc\hspace{-0.3em}\text{---} \;\; := \sum_{i,j} \;\text{---}\bigcirc\hspace{-0.3em}{}^{\,i}_{\,j} \quad {}^{\,i}_{\,j}\bigcirc\hspace{-0.3em}\text{---} \quad (1.7)$$

Familiar examples of contraction are vector inner products, matrix-vector multiplication, matrix-matrix multiplication, and the trace of a matrix:

| Conventional | Einstein | TNN |
|:---:|:---:|:---:|
| $\langle \vec{x}, \vec{y} \rangle$ | $x_\alpha y^\alpha$ | $\boxed{x}\text{---}\boxed{y}$ |
| $M\vec{v}$ | $M^\alpha_{\ \beta} v^\beta$ | $\text{---}\boxed{M}\text{---}\boxed{v}$ |
| $AB$ | $A^\alpha_{\ \beta} B^\beta_{\ \gamma}$ | $\text{---}\boxed{A}\text{---}\boxed{B}\text{---}$ |
| $\text{Tr}(X)$ | $X^\alpha_{\ \alpha}$ | $\boxed{X}$ |

#### 1.2.4 Grouping and splitting

Rank is a rather fluid concept in the study of tensor networks. The space of tensors $\mathbb{C}^{a_1 \times \cdots \times a_n}$ and $\mathbb{C}^{b_1 \times \cdots \times b_m}$ are isomorphic as vector spaces whenever the overall dimensions match ($\prod_i a_i = \prod_i b_i$). Using this we can extend concepts and techniques only previously defined for vectors and matrices to all tensors. To do this, we can *group* or *split* indices to lower or raise the rank of a given tensor respectively.

Consider the case of contracting two arbitrary tensors. If we group together the indices which are and are not involved in this contraction, this procedure simply reduces to matrix multiplication:

$$\raisebox{-0.5em}{\includegraphics{eq18}} \quad (1.8)$$

It should be noted that not only is this reduction to matrix multiplication pedagogically handy, but this is precisely the manner in which numerical tensor packages perform contraction, allowing them to leverage highly optimised matrix multiplication code.

Owing to the freedom in choice of basis, the precise details of grouping and splitting are not unique. One specific choice of convention is the *tensor product basis*, defining a basis on the product space simply given by the product of the respective bases. The canonical use of tensor product bases in quantum information allows for the grouping and splitting described above to be dealt with implicitly. Statements such as $|0\rangle \otimes |1\rangle \equiv |01\rangle$ omit precisely this grouping: notice that the tensor product on the left is a $2 \times 2$ dimensional matrix, whilst the right hand-side is a 4-dimensional vector. The 'tensor product' used in quantum information is often in fact a *Kronecker product*, given by a true tensor product followed by just such a grouping.

More concretely, suppose we use an index convention that can be considered a higher-dimensional generalisation of column-major ordering. If we take a rank $n + m$ tensor, and group its first $n$ indices and last $m$ indices together to form a matrix

$$T_{I,J} := T_{i_1,\ldots,i_n;j_1,\ldots,j_m} \quad (1.9)$$

where we have defined our grouped indices as

$$I := i_1 + d_1^{(i)} \cdot i_2 + d_1^{(i)} d_2^{(i)} \cdot i_3 + \cdots + d_1^{(i)} \ldots d_{n-1}^{(i)} \cdot i_n, \quad (1.10)$$

$$J := j_1 + d_1^{(j)} \cdot j_2 + d_1^{(j)} d_2^{(j)} \cdot j_3 + \cdots + d_1^{(j)} \ldots d_{m-1}^{(j)} \cdot j_m, \quad (1.11)$$

where $d_x^{(i)}(d_x^{(j)})$ is the dimension of the $x$th index of type $i(j)$. When such a grouping is given, we can now treat this tensor as a matrix, performing standard matrix operations.

An important example is the singular value decomposition (SVD), given by $T_{I,J} = \sum_\alpha U_{I,\alpha} S_{\alpha,\alpha} \bar{V}_{J,\alpha}$. By performing the above grouping, followed by the SVD, and then splitting the indices back out, we get a higher dimensional version of the SVD

$$T_{i_1,\ldots,i_n;j_1,\ldots,j_m} = \sum_\alpha U_{i_1,\ldots,i_n,\alpha} S_{\alpha,\alpha} \bar{V}_{j_1,\ldots,j_m,\alpha}.$$

So long as we choose them to be consistent, the precise method by which we group and split is immaterial in this overall operation. As a result we will keep this grouping purely implicit, as in the first equality Eq. (1.8). This will be especially useful for employing notions defined for matrices and vectors to higher rank objects, implicitly grouping then splitting. Graphically the above SVD will simply be denoted

$$\raisebox{-0.5em}{\includegraphics{eq112}} \quad , \quad (1.12)$$

where $U$ and $V$ are isometric ($U^\dagger U = V^\dagger V = \mathbb{1}$) across the indicated partitioning, and where the conjugation in $V^\dagger$ is included for consistency with conventional notation and also taken with respect to this partitioning. We will refer to such a partitioning of the indices in to two disjoint sets as a *bisection* of the tensor.

## 1.3 Tensor networks

Combining the above tensor operations, we can now give a single definition of a tensor network. A tensor network is a diagram which tells us how to combine several tensors into a single composite tensor. The rank of this overall tensor is given by the number of unmatched legs in the diagram. The value for a given configuration of external indices, is given by the product of the values of the constituent tensors, summed over all internal index labellings consistent with the contractions. A generic example of this is given below:

$$\text{(1.13)}$$

## 1.4 Bubbling

Whilst tensor networks are defined in such a way that their values are independent of the order in which the constituent tensors are contracted, such considerations do influence the complexity and practicality of such computations. Tensor networks can be contracted by beginning with a single tensor and repeatedly contracting it against tensors one-at-a-time. The order in which tensors are introduced and contracted is known as a *bubbling*. As the bubbling is performed the network is swallowed into the stored tensor, until only the result remains.

Many networks admit both efficient and inefficient bubblings, highlighting the need for prudence when planning out contractions. Take for example a ladder-shaped network (we'll see a few of these in the following lectures). One bubbling we may consider is to contract along the top of the ladder, then back along the bottom. Showing both this bubbling, as well as the partially contracted tensor that is kept in memory (in red), we see this bubbling looks like:

$$\text{(1.14)}$$

$$\text{(1.15)}$$

The scaling of this procedure is however quite unfavourable; consider a ladder of length $n$. At the midpoint of this contraction, when the top has been contracted, the tensor being tracked has rank $n$, and thus the number of entries is scaling exponentially with $n$. As such the memory and time footprints of this contraction are also exponential, rendering it infeasible for large $n$. If however we contract each rung in turn, the tracked tensor has a rank never more than 3, giving constant memory and linear time costs.

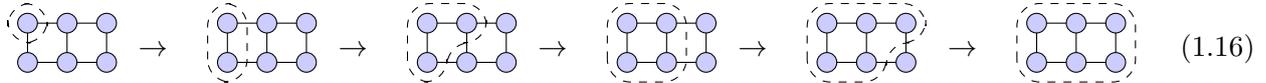 (1.16)

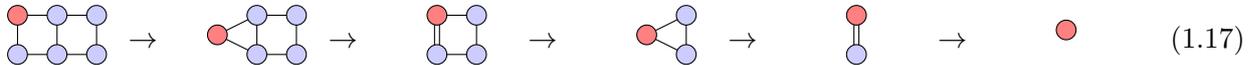 (1.17)

The memory footprint at any step during the contraction corresponds to the product of the dimensions of each leg passing through the boundary of the contracted region (see the red legs in Eq. (1.18)). Whilst the above ladder arrangement possesses both good and bad bubblings, some networks possess an underlying graph structure that does not admit *any* efficient contraction ordering. A good example of this is the 2D grid; due to the 2D structure of this lattice, it is clear that the contracted region must, somewhere near the middle of the contracting procedure, have a perimeter on the order of $\sqrt{n}$ where $n$ is the number of tensors. As a result such contractions generically take exponential time/memory to perform. An example of a high cost step during such a bubbling is shown below, with the prohibitively large perimeter indicated by the red legs.

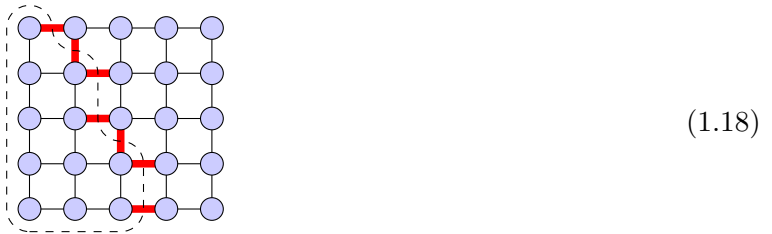 (1.18)

Although the bubblings we have depicted here involve picking a single tensor and contracting others into it one-by-one, this will frequently not be the most efficient order; often a multibubbling approach is faster. Ref. [3] provides code which allows for finding optimal bubbling order for networks of up to 30-40 tensors. This code interfaces with that provided in Ref. [4] and Ref. [5], providing a complete tensor network package.

## 1.5 Computational Complexity

Above we've described that there exist networks which stymie the specific contraction procedures we've outlined. In this section we'll see that there also exist networks for which there are complexity theoretic obstructions which do not allow for *any* contraction procedure to be efficient.

We will now consider the computational complexity associated with tensor network contractions. Whilst all of the tensor networks we will consider in later lectures constitute memory-efficient representations of objects such as quantum states, not all permit efficient manipulation. This demonstrates that how one wishes to manipulate a tensor network is an important part of considering them as ansätze.

Whilst algorithms which can speed up tensor network contractions by optimising the bubbling used [3–5], as discusssed above, the underlying computational problem is NP-complete [6,7]

Even ignoring the specific bubbling used, the complexity of the overall contraction procedure can also be shown to be prohibitive in general. Consider a network made from the binary tensors $e$ and $n$. The value of $e$ is 1 if and only if all indices are identical, and zero otherwise, whilst $n$ has value 1 if and only if all legs differ and 0 otherwise. Take an arbitrary graph, and construct a tensor network with an $e$ tensor at each vertex and $n$ tensor in the middle of each edge, with the connectedness inherited from the graph.

$$\longrightarrow \qquad\qquad (1.19)$$

By construction, the non-zero contributions to the above tensor network correspond to an assignment of index values to each vertex (enforced by $e$) of the original graph, such that no two neighbouring vertices share the same value (enforced by $n$). If each index is $q$-dimensional this is a vertex $q$-colouring of the graph, and the value of the tensor network corresponds to the number of such $q$-colourings. As determining the existence of a $q$-colouring is an NP-complete problem [8], contracting this graph is therefore #P-complete [9]. Indeed similar constructions exist for tensor networks corresponding to #SAT and other #P-complete problems [10]. As we will see later in Section 6, there also exists a quantum hardness result which shows approximate contraction to be Post-BQP-hard, putting it inside a class of problems not believed to be efficiently solvable on even a quantum computer.

---

## Problems 1

Solutions in accompanying document.

1. Consider the following tensors, in which all indices are three-dimensional, indexed from 0:

$$\begin{array}{cc}
\underset{j}{\overset{i}{A}} = i^2 - 2j, & \overset{i}{\underset{k}{B}}{}^{j} = -3^i j + k, \quad (1.20)
\end{array}$$

$$\begin{array}{cc}
i-\overset{j}{C} = j & \overset{i}{\underset{j}{\phantom{x}}}\!D\overset{}{\underset{k}{\phantom{x}}} = ijk. \quad (1.21)
\end{array}$$

Calculate the value of the following tensor network:

$$\begin{array}{c}
A - D \\
| \quad | \\
B - C
\end{array} \qquad\qquad (1.22)$$

2. In this question we are going to consider expanding out a contraction sequence, in a manner which would be needed when coding up contractions. Given a network, and an associated bubbling, we wish to write out a table keeping track of the indices of the current object, the tensor currently being contracted in, the indices involved in that contraction, and new indices left uncontracted. For example for the network

$$\begin{array}{c}
\overset{\alpha}{\underset{\beta}{A}}\overset{B}{\phantom{x}}\overset{\gamma}{\underset{}{C}}-\delta
\end{array} \qquad\qquad (1.23)$$

where the bubbling is performed in alphabetical order, then the table in question looks like

| Current | Tensor | Contract | New |
|---------|--------|----------|-----|
| – | $A$ | – | $\alpha, \beta$ |
| $\alpha, \beta$ | $B$ | $\alpha$ | $\gamma$ |
| $\beta, \gamma$ | $C$ | $\beta, \gamma$ | $\delta$ |

For the tensor network

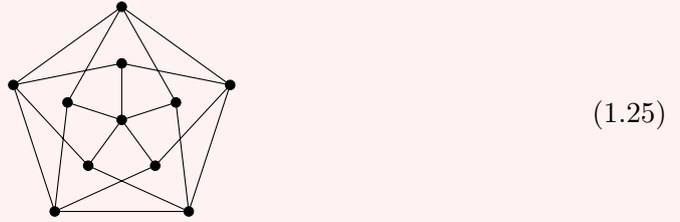

$$\tag{1.24}$$

construct a corresponding table, where contraction is once again done in alphabetical order.

3.  (a) Calculate the contraction of the tensor network in Eq. (1.19) for bond dimension 3, i.e. calculate the number of three-colourings of the corresponding graph.

    (b) Using the $e$ and $n$ tensors from Section 1.5, come up with a construction for a tensor network which gives the number of *edge* colourings. For a planar graphs, construct an analogous network to count *face* colourings.

    (c) Using tensor networks, determine the minimum number of colours required to vertex and edge colour the below graph (known as the chromatic number and index respectively).

$$\tag{1.25}$$

4.  Much like the singular value decomposition, given a bisection of the indices we can consider norms of tensors.

    (a) Does the operator norm depend on the bisection, i.e. are the operator norms across any two bisections of the same tensor necessarily equal?

    (b) What about the Frobenius norm? If they can differ, give an example, if not draw a tensor network diagram that shows it to be manifestly independent of bisection.

5.  Write out the Einstein notation corresponding to the network in Eq. (7.1).

## 1.7 References

[1] C. Eckart and G. Young, "The approximation of one matrix by another of lower rank," *Psychometrika* **1**, (1936).

[2] L. Mirsky, "Symmetric gauge functions and unitarily invariant norms," *The Quarterly Journal of Mathematics* **11**, 1, 50–59, (1960).

[3] R. N. C. Pfeifer, J. Haegeman, and F. Verstraete, "Faster identification of optimal contraction sequences for tensor networks," *Physical Review E* **90** 033315, arXiv:1304.6112, (2014).

[4] R. N. C. Pfeifer, G. Evenbly, S. Singh, and G. Vidal, "NCON: A tensor network contractor for MATLAB," arXiv:1402.0939v1, (2014).

[5] G. Evenbly and R. N. C. Pfeifer, "Improving the efficiency of variational tensor network algorithms," *Physical Review B* **89** 245118, arXiv:1402.0939v1, (2014).

[6] I. Arad and Z. Landau, "Quantum computation and the evaluation of tensor networks," *SIAM Journal on Computing* **39** 3089, arXiv:0805.0040, (2010).

[7] L. Chi-Chung, P. Sadayappan, and R. Wenger, "On Optimizing a Class of Multi-Dimensional Loops with Reduction for Parallel Execution," *Parallel Processing Letters* **07** 157–168, (1997).

[8] M. R. Garey, D. S. Johnson, and L. Stockmeyer, "Some simplified NP-complete problems," in *Proceedings of the sixth annual ACM symposium on Theory of computing - STOC '74*, 47–63, ACM Press, (1974).

[9] M. Dyer, L. A. Goldberg, C. Greenhill, and M. Jerrum, "The Relative Complexity of Approximate Counting Problems," *Algorithmica* **38** 471–500, (2004).

[10] J. D. Biamonte, J. Morton, and J. Turner, "Tensor Network Contractions for #SAT," *Journal of Statistical Physics* **160** 1389–1404, arXiv:1405.7375, (2015).

## 2 Quantum information examples

In this lecture we will cover a few examples of concepts in quantum information which can be better understood in tensor network notation. This lecture won't serve as much as an introduction to these concepts, but instead as a Rosetta stone for those familiar with quantum information and not with TNN. For a more thorough introduction to quantum information see the textbooks of Refs. [1–3] or lecture notes of Refs. [4,5]. We note that for the study of open quantum systems, a more specialised form of TNN was developed in Ref. [6].

### 2.1 Bell state and the Bell basis

The Bell basis forms a convenient orthonormal set of two qubit states that exhibit maximal entanglement. The standard notation for this basis is

$$|\Phi^\pm\rangle := \Big(|0\rangle \otimes |0\rangle \pm |1\rangle \otimes |1\rangle\Big)/\sqrt{2} \qquad \text{and} \qquad |\Psi^\pm\rangle := \Big(|0\rangle \otimes |1\rangle \pm |1\rangle \otimes |0\rangle\Big)/\sqrt{2}.$$

The first of this basis, $|\Phi^+\rangle$, we shall denote $|\Omega\rangle$ and simply refer to as *the* Bell state. Thought of as a matrix, $\Omega$ is proportional to the one qubit identity,

$$|\Omega\rangle = \frac{1}{\sqrt{2}} \begin{pmatrix} 1 \\ 0 \\ 0 \\ 1 \end{pmatrix} \quad \xleftrightarrow[\text{Matricise}]{\text{Vectorise}} \quad \frac{1}{\sqrt{2}} \begin{pmatrix} 1 & 0 \\ 0 & 1 \end{pmatrix} = I/\sqrt{2}. \tag{2.1}$$

In tensor network notation, this is represented simply as a line connecting its two legs.

$$\boxed{\Omega}\;\; = \frac{1}{\sqrt{2}} \;\;\supset \tag{2.2}$$

Next we will define $\Omega(O)$ to be the vectorisation of an operator $O$, such that $|\Omega(O)\rangle = (O \otimes I)|\Omega\rangle$.

$$\boxed{\Omega(O)}\;\; = \frac{1}{\sqrt{2}} \;\;\boxed{O}\supset \tag{2.3}$$

Given this definition, we can see that the Bell basis simply corresponds to a vectorisation of the Pauli operators

$$|\Phi^+\rangle = |\Omega(I)\rangle, \quad |\Phi^-\rangle = |\Omega(Z)\rangle, \quad |\Psi^+\rangle = |\Omega(X)\rangle, \quad |\Psi^-\rangle \propto |\Omega(Y)\rangle.$$

Thus we see that the Bell basis is intimately linked to the Pauli operators, with the Euclidean inner product on Bell basis states corresponding to the Hilbert-Schmidt inner product on Paulis.

### 2.2 Quantum Teleportation

Given this notation for the Bell basis, we can now understand Quantum Teleportation in TNN. The idea here is for two parties (Alice and Bob, say) to share a Bell state. Given this shared resource of entanglement, we then allow Alice to perform local operations on her half of the pair, and an arbitrary fiducial qubit. After transmitting only two classical bits, Bob can then correct his half of the pair such that he recovers the state of the original fiducial qubit, successfully teleporting the data within.

The procedure for teleportation goes as follows. First Alice performs a projective measurement in the Bell basis on both the fiducial qubit and her Bell qubit, receiving outcome $|\Omega(p)\rangle$. The result of this measurement is then (classically) transmitted to Bob, requiring two communication bits. Bob then performs the corresponding Pauli $p$ on his Bell qubit, correcting the influence of the measurement. Taking the fiducial state to be $|\psi\rangle$, and supposing the measurement corresponds to $|\Omega(p)\rangle$, then this procedure gives Bob a final state of $|\phi\rangle = |\psi\rangle/2$:

$$|\phi\rangle = \overbrace{(p_B}^{\text{Correction}} \overbrace{\langle\Omega_{A_1 A_2}(p)|)}^{\text{Teleportation}} \overbrace{(|\psi_{A_1}\rangle \otimes |\Omega_{A_2 B}\rangle)}^{\text{Setup}} = |\psi\rangle/2 \qquad (2.4)$$

where $A_1$ and $A_2$ correspond to the single qubit registers of Alice, and $B$ to Bob's qubit. In tensor network notation this can be clearly seen:

$$\qquad (2.5)$$

$$= \frac{1}{2} \quad \boxed{p}\!-\!\boxed{p^\dagger}\!-\!\boxed{\psi} \qquad (2.6)$$

$$= |\psi\rangle/2 \qquad (2.7)$$

where the dashed line indicates the physical separation of the two parties.

As such we can see that $|\psi\rangle$ is correctly transmitted for any measurement outcome $p$, each of which is seen with probability $1/4$. Thus we see that in spite of the non-deterministic intermediary states, the overall procedure is deterministic. Analogous procedures can work for $p$ being elements of any set of operators which are orthonormal with respect to the Hilbert-Schmidt inner product, e.g. higher dimensional Paulis.

### 2.2.1 Gate Teleportation

The idea behind gate teleportation is similar to regular teleportation, but utilises a general maximally entangled state instead of the Bell state specifically. Suppose we prepare a maximally entangled state $|\Omega(U^T)\rangle$ corresponding to a unitary $U$, and post select on a Bell basis measurement of $|\Omega(p)\rangle$, followed by a correcting unitary $C_p$, then Bob ends up with the state:

$$|\phi\rangle = \overbrace{(C_p}^{\text{Correction}} \overbrace{\langle\Omega_{A_1 A_2}(p)|)}^{\text{Teleportation}} \overbrace{(|\psi_{A_1}\rangle \otimes |\Omega_{A_2 B}(U^T)\rangle)}^{\text{Setup}} \qquad (2.8)$$

$$\qquad (2.9)$$

$$= \frac{1}{2} \quad \boxed{C_p}\!-\!\boxed{U}\!-\!\boxed{p^\dagger}\!-\!\boxed{\psi} \qquad (2.10)$$

$$= C_p U p^\dagger |\psi\rangle/2 \qquad (2.11)$$

If we take $C_p := U p U^\dagger$ then Bob receives $U|\psi\rangle$ for all measurement outcomes, i.e. $|\phi\rangle \propto U|\psi\rangle$. If $U$ is a Clifford operator[2], this correction is also a Pauli, making the procedure no more resource intensive in terms of the gates used than standard teleportation.

An example of where this is useful is in the case where Paulis can be reliably performed, but Cliffords can only be applied non-deterministically. Gate teleportation allows us to prepare the $|U^T\rangle$ first, simply retrying the non-deterministic procedure until it succeeds. Once this has succeeded, we can use gate teleportation to apply this unitary on the data state using only Pauli operations. As such we can avoid needing to apply non-deterministic gates directly on our target state, endangering the data stored within.

---

[2]The Cliffords are the group of unitaries which map Paulis to Paulis under conjugation.

## 2.3 Purification

For a given mixed state $\rho$, a purification is a pure state $|\psi\rangle$ which is extended into a larger system (the added subsystem is known as the *purification system*), such that the reduced density on the original system is $\rho$. One such purification is given by $|\psi\rangle \propto (\sqrt{\rho} \otimes I)|\Omega\rangle = |\Omega(\sqrt{\rho})\rangle$, which can be simply seen by considering the corresponding tensor networks. The definition of the state is

$$\psi = \sqrt{\rho} \qquad (2.12)$$

which gives a reduced density of

$$\mathrm{Tr}_2\Big(|\psi\rangle\langle\psi|\Big) = \psi \ \psi = \sqrt{\rho} \ \sqrt{\rho} = \rho \qquad (2.13)$$

By dimension counting, it can be shown that the above purification is unique up to an isometric freedom on the purification system, i.e. all purifications are of the form $\big(\sqrt{\rho} \otimes U\big)|\Omega\rangle$ where $U^\dagger U = \mathbb{1}$. Equivalently all purifications can be considered to be proportional to $(\sqrt{\rho} \otimes I)|\Omega\rangle$, where $|\Omega\rangle$ is some maximally entangled state other than the Bell state.

## 2.4 Stinespring's Dilation Theorem

Stinespring's Theorem says that any quantum channel $\mathcal{E}$ – a completely positive trace preserving (CPTP) map – can be expressed as a unitary map $V$ acting on a larger system followed by a partial trace, i.e.

$$\mathcal{E}(\rho) = \mathrm{Tr}_1\Big[V^\dagger\,(\rho \otimes |0\rangle\langle 0|)\,V\Big]. \qquad (2.14)$$

Physically this means that dynamics of an open system is equivalent to those of a subsystem of a larger, closed system — the founding tennet of the Church of the Larger Hilbert Space. Any CPTP map can be represented by a set of Kraus operators $K_i$ such that

$$\mathcal{E}(\rho) = \sum_i K_i^\dagger \rho K_i \ \text{ where } \ \sum_i K_i K_i^\dagger = I. \qquad (2.15)$$

In TNN this looks like

$$\mathcal{E} \ \rho = K^\dagger \ \rho \ K \quad \text{where} \quad K \ K^\dagger = \qquad (2.16)$$

where the transposition in the Hermitian conjugate is done with respect to the horizontal legs, and the upper leg corresponds to the virtual index $i$.

Next we define the tensor $U$ as

$$U := K^\dagger \qquad (2.17)$$

where we can see that $U$ is an isometry ($U^\dagger U = I$), which we can think of as a unitary $V$ with an omitted ancilla

$$U = V\,|0\rangle \quad . \qquad (2.18)$$

Using this, and partial tracing over the upper index, we get the Stinespring Dilation Theorem as desired:

$$\mathcal{E}(\rho) = \sum_i K_i^\dagger \rho K_i = \quad \boxed{K^\dagger}\,\boxed{\rho}\,\boxed{K} \tag{2.19}$$

$$= \quad \boxed{K^\dagger}\,\boxed{\rho}\,\boxed{K} \tag{2.20}$$

$$= \quad \boxed{U}\,\boxed{\rho}\,\boxed{U^\dagger} \tag{2.21}$$

$$= \quad \boxed{V}\,\boxed{|0\rangle\langle0|}{\atop\boxed{\rho}}\,\boxed{V^\dagger} \tag{2.22}$$

$$= \mathrm{Tr}_1\left[V^\dagger\left(\rho \otimes |0\rangle\langle0|\right)V\right] \tag{2.23}$$

---

### Problems 2

Solutions in accompanying document.

1. Consider the inverse of teleportation. Alice wishes to send classical bits to Bob, and possesses a quantum channel through which she can send Bob qubits. How many bits of information can be communicated in a single qubit? For simplicity consider the case where Bob can only perform projective measurements.

2. Suppose Alice and Bob initially shared a Bell pair. Does this pre-shared entanglement resource boost the amount of classical information that can be successfully communicated, and if so by how much? *Hint: Notice that the four possible Bell states differ by a Pauli acting on a single qubit.*

## 2.6 References

[1] M. A. Nielsen and I. L. Chuang, Quantum Computation and Quantum Information 10th Anniversary Edition, Cambridge University Press, (2011).

[2] N. D. Mermin, Quantum Computer Science: An Introduction, Cambridge University Press, (2007), http://www.lassp.cornell.edu/mermin/qcomp/CS483.html.

[3] M. M. Wilde, Quantum Information Theory, Cambridge University Press, arXiv:1106.1445 (2013).

[4] J. Preskill, Quantum Computation, http://www.theory.caltech.edu/people/preskill/ph229/.

[5] J. Watrous, Theory of Quantum Information and Introduction to Quantum Computing, https://cs.uwaterloo.ca/ watrous/LectureNotes.html.

[6] C. J. Wood, J. D. Biamonte, and D. G. Cory, "Tensor networks and graphical calculus for open quantum systems," *Quantum Information and Computation* **15**, 0759, arXiv:1111.6950, (2011).

# 3 Matrix Product States

Now that we have established the notation, the remaining lectures will examine some key tensor networks and algorithms for strongly interacting quantum many body systems. We begin with one dimensional models.

Matrix product states (MPS) are a natural choice for efficient representation of 1D quantum low energy states of physically realistic systems [1–6]. This lecture will begin by motivating and defining MPS in two slightly different ways. We will then give some analytic examples of MPS, demonstrating some of the complexity which can be captured with this simple network. Some simple properties of MPS will then be explained, followed by a generalisation of the network to operators rather than pure states.

Let $|\psi\rangle = \sum_{j_1 j_2 \dots j_N = 0}^{d-1} C_{j_1 j_2 \dots j_N} |j_1\rangle \otimes |j_2\rangle \otimes \cdots \otimes |j_N\rangle$ be the (completely general) state of $N$ qudits (d dimensional quantum system). The state is completely specified by knowledge of the rank-$N$ tensor $C$.

By splitting the first index out from the rest, and performing an SVD, we get the Schmidt decomposition

$$|\psi\rangle = \sum_i \lambda_i |L_i\rangle \otimes |R_i\rangle, \tag{3.1}$$

where $\lambda_i$ are the Schmidt weights and $\{|L_i\rangle\}$ and $\{|R_i\rangle\}$ are orthonormal sets of vectors. Graphically this looks like

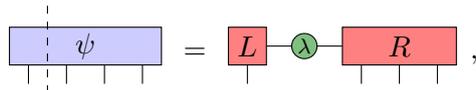

$$\tag{3.2}$$

where $\lambda$ is a diagonal matrix containing the Schmidt weights.

The $\alpha$-Rényi entropy is given by

$$S_\alpha(\rho) = \frac{1}{1-\alpha} \log \operatorname{Tr} \rho^\alpha, \tag{3.3}$$

where $\rho$ is some density matrix. Note that the *entanglement rank* $S_0$ is simply the (log of the) number of nonzero Schmidt weights and the von Neumann entropy is recovered for $\alpha \to 1$. We also note that the Schmidt weights now correspond precisely to the singular values of the decomposition Eq. (3.2), and so these values capture the entanglement structure along this cut.

We can now perform successive singular value decompositions along each cut in turn, splitting out the tensor into local tensors $M$, and diagonal matrices of singular values $\lambda$ quantifying the entanglement across that cut.

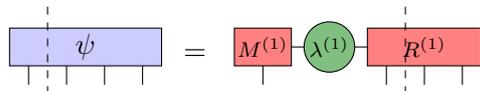

$$\tag{3.4}$$

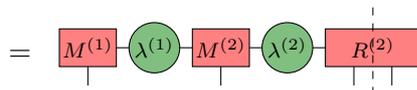

$$\tag{3.5}$$

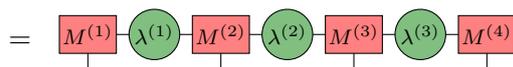

$$\tag{3.6}$$

By now contracting[3] the singular values tensors $\lambda^{(i)}$ into the local tensors $M^{(i)}$ we get the more generic form

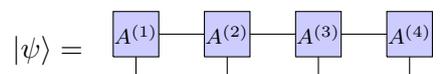

$$\tag{3.7}$$

---

[3]Into precisely which tensor the singular values are contracted can be important, and relates to gauge fixing the MPS, see Section 3.3.2.

This is the *matrix product state*. It is not yet clear that we have done anything useful. The above construction is both general and exact, so we have the same number of coefficients in an arguably much more complicated form.

Suppose however we consider states for which the entanglement rank across any bisection of the chain is bounded. In particular, suppose that only $D$ of the Schmidt weights were non-zero. Then we can use the MPS form to take advantage of this by truncating the $\lambda$ matrix to make use of this property. In particular, any state with a so-called *strong area law* such that $S_0 \leq \log c$ for some constant $c$ along any bipartition can be expressed (exactly) using an MPS with only $\mathcal{O}(dNc^2)$ coefficients. As discussed in Sec. 5, there are many relevant states for which an area law for the von Neumann entropy ($S_1 = \mathcal{O}(1)$) is sufficient to guarantee arbitrarily good approximation with an MPS of only $\mathsf{poly}(N)$ bond dimension [1–3].

In TNN, the name matrix product state is a misnomer, as most tensors involved are in fact rank-3. The uncontracted index is referred to as the *physical* index, whilst the other two are *virtual*, *bond* or *matrix* indices. For reasons of convenience, as well as to capture periodic states most efficiently, the MPS ansatz is usually modified from Eq. (3.7) to

$$\left| \psi \left[ A^{(1)}, A^{(2)}, \ldots, A^{(N)} \right] \right\rangle = \sum_{i_1 i_2 \ldots i_N} \mathrm{Tr} \left[ A_{i_1}^{(1)} A_{i_2}^{(2)} \ldots A_{i_N}^{(N)} \right] |i_1 i_2 \ldots i_N\rangle, \tag{3.8}$$

or in the translationally invariant case

$$|\psi[A]\rangle = \sum_{i_1 i_2 \ldots i_N} \mathrm{Tr} \left[ A_{i_1} A_{i_2} \ldots A_{i_N} \right] |i_1 i_2 \ldots i_N\rangle. \tag{3.9}$$

Note that in this form the matrix indices are suppressed and matrix multiplication is implied. The graphical form of this MPS is

$$|\psi[A]\rangle = \quad \text{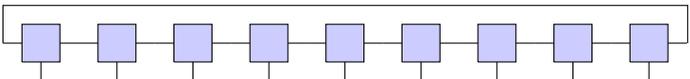} \tag{3.10}$$

## 3.1 1D Projected Entangled Pair States

In addition to the above construction, MPS can (equivalently) be viewed as a special case of the *projected entangled pair states* (PEPS) construction [2, 7, 8]. This proceeds by laying out entangled pair states $|\phi\rangle$ on some lattice and applying some linear map $\mathcal{P}$ between pairs

$$|\psi\rangle = \quad \text{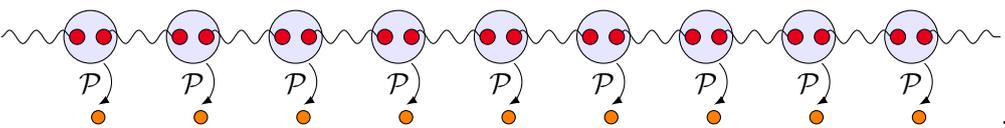} \tag{3.11}$$

where

$$|\phi\rangle = \quad \bullet\!\!\sim\!\!\sim\!\!\sim\!\!\bullet \tag{3.12}$$

is the chosen entangled pair. In Lecture 6, we will generalise this construction to arbitrary dimensions and arbitrary lattices.

It is clear that this construction is equivalent to the tensor network construction by letting $|\phi\rangle = \sum_{j=0}^{d-1} |dd\rangle$. We can write the linear map $\mathcal{P}$ as

$$\mathcal{P} = \sum_{i,\alpha,\beta} A_{i;\alpha,\beta} |i\rangle\langle\alpha\beta|. \tag{3.13}$$

The tensor $A$ is exactly the MPS tensor introduced above, and the choice of entangled pair ensures that the $A$ tensor corresponding to a pair of PEPS 'projectors' applied to the Bell state above is exactly the contraction of the corresponding $A$ tensors:

$$\mathcal{P}^{(1)} \otimes \mathcal{P}^{(2)} |\phi\rangle_{2,3} = \sum_{i_1,i_2,\alpha_1,\beta_1,\alpha_2,\beta_2,j} A_{i_1;\alpha_1,\beta_1}^{(1)} A_{i_2;\alpha_2,\beta_2}^{(2)} |i_1 i_2\rangle\langle\alpha_1\beta_1\alpha_2\beta_2|(\mathbb{1} \otimes |jj\rangle \otimes \mathbb{1}) \tag{3.14}$$

$$= \sum_{i_1,i_2;\alpha_1,\beta_1,\beta_2} A^{(1)}_{i_1;\alpha_1,\beta_1} A^{(2)}_{i_2;\beta_1,\beta_2} |i_1 i_2\rangle\langle\alpha_1\beta_2|. \tag{3.15}$$

Thus, we see that the two descriptions are equivalent, and interchanged through the applications of local unitaries to the virtual indices of $A$ or equivalently changing the maximally entangled pair in the PEPS.

We note that this should not generally be seen as a practical preparation procedure. Generically the PEPS tensors will map states down into a non-trivial subspace, with the physical implementation of this requiring post-selected measurements. If one of these fails, we need to go back and begin the construction from the start, meaning this procedure is not generally scalable.

## 3.2 Some MPS states

**Product State**

Let

$$A_0 = \begin{pmatrix} 1 \end{pmatrix}, \qquad\qquad A_1 = \begin{pmatrix} 0 \end{pmatrix}. \tag{3.16}$$

This gives the state $|00\ldots0\rangle$, as does

$$A_0 = \begin{pmatrix} 1 & 0 \\ 0 & 0 \end{pmatrix}, \qquad\qquad A_1 = \begin{pmatrix} 0 & 0 \\ 0 & 0 \end{pmatrix}. \tag{3.17}$$

**W State**

What state do we get when we set

$$A_0 = \begin{pmatrix} 1 & 0 \\ 0 & 1 \end{pmatrix}, \qquad\qquad A_1 = \begin{pmatrix} 0 & 1 \\ 0 & 0 \end{pmatrix}, \tag{3.18}$$

and we choose the boundary conditions of the MPS to be

$$|\psi[A]\rangle = \boxed{\phantom{xx}\,\color{red}{X}} \quad ? \tag{3.19}$$

We have $A_0 A_0 = A_0$, $A_0 A_1 = A_1$, $A_1^2 = 0$ and $\mathrm{Tr}[A_1 X] = 1$, so we get

$$|W\rangle = \sum_{j=1}^{N} |000\ldots01_j000\ldots0\rangle, \tag{3.20}$$

the W-state [2].

**GHZ State**

If we choose $|\phi\rangle = |00\rangle + |11\rangle$ and $\mathcal{P} = |0\rangle\langle00| + |1\rangle\langle11|$, or the equivalent MPS tensor

$$A_0 = \begin{pmatrix} 1 & 0 \\ 0 & 0 \end{pmatrix}, \qquad\qquad A_1 = \begin{pmatrix} 0 & 0 \\ 0 & 1 \end{pmatrix}, \tag{3.21}$$

then we get the Greenberger-Horne-Zeilinger (GHZ) state [2]

$$|GHZ\rangle = |00\ldots0\rangle + |11\ldots1\rangle. \tag{3.22}$$

**AKLT State**

Suppose we wish to construct an $SO(3)$ symmetric spin-1 state [5, 6, 9]. Let $|\phi\rangle = |01\rangle - |10\rangle$ be the $SO(3)$ invariant singlet state. Let $\mathcal{P} : \mathbb{C}^{2\times 2} \to \mathbb{C}^3$ be the projector onto the spin-1 subspace

$$\mathcal{P} = |\bar{1}\rangle\langle 00| + |\bar{0}\rangle \frac{\langle 01| + \langle 10|}{\sqrt{2}} + |-\bar{1}\rangle\langle 11|. \tag{3.23}$$

The advantage is that the spin operators on the corresponding systems pull through $\mathcal{P}$, meaning it commutes with rotations. Let $(S_x, S_y, S_z)$ be the spin vector on the spin-1 particle, and $(X_i, Y_i, Z_i)/2$ the spin vector on the $i$th qubit, then this means:

$$S_Z\mathcal{P} = \left(|\bar{1}\rangle\langle\bar{1}| - |-\bar{1}\rangle\langle -\bar{1}|\right)\left(|\bar{1}\rangle\langle 00| + |\bar{0}\rangle \frac{\langle 01| + \langle 10|}{\sqrt{2}} + |-\bar{1}\rangle\langle 11|\right) \tag{3.24}$$

$$= |\bar{1}\rangle\langle 00| - |-\bar{1}\rangle\langle 11| \tag{3.25}$$

$$= \mathcal{P}\frac{Z_1 + Z_2}{2} \tag{3.26}$$

$$S_X\mathcal{P} = \frac{|\bar{0}\rangle\left(\langle\bar{1}| + \langle-\bar{1}|\right) + \left(|\bar{1}\rangle + |-\bar{1}\rangle\right)\langle\bar{0}|}{\sqrt{2}}\left(|\bar{1}\rangle\langle 00| + |\bar{0}\rangle \frac{\langle 01| + \langle 10|}{\sqrt{2}} + |-\bar{1}\rangle\langle 11|\right) \tag{3.27}$$

$$= \left(\frac{|\bar{0}\rangle\left(\langle 00| + \langle 11|\right)}{\sqrt{2}} + \frac{\left(|\bar{1}\rangle + |-\bar{1}\rangle\right)\left(\langle 01| + \langle 10|\right)}{2}\right) \tag{3.28}$$

$$= \mathcal{P}\frac{X_1 + X_2}{2}, \tag{3.29}$$

with the same holding for $S_Y$. Thus the state obtained after this projection is fully $SO(3)$ symmetric, but has a nontrivial entanglement structure (which would not be obtained if the state was simply a singlet at each site for example).

This state has many interesting properties. We can write a 2-local Hamiltonian for which this is the ground state. Let $\Pi_2$ be the projector onto the spin-2 subspace of a pair of spin-1 particles. This operator has eigenvalues $\{0, 1\}$. $\Pi_2$ annihilates an adjacent pair of spin-1 particles, since they are built from two spin-1/2s and a spin-0, so have no overlap with the spin-2 subspace. It is simple to check that on periodic boundary conditions the ground state of $H = \sum \Pi_2$ is unique (and gapped).

If we examine the action of rotations about the three axes of the spin-1, we see that

$$R_{\hat{n}}(\theta)\mathcal{P} = \mathcal{P}R_{\hat{n}}(\theta) \otimes R_{\hat{n}}(\theta). \tag{3.30}$$

In particular, $R_{\hat{x}}(\pi) \mapsto -XX$, $R_{\hat{y}}(\pi) \mapsto -YY$, $R_{\hat{z}}(\pi) \mapsto -ZZ$. In Sec. 4 we will see that this tells us the AKLT state is in a nontrivial symmetry protected topological (SPT) phase.

**Cluster State**

It is convenient to write a bond dimension 2 MPS for this state where a physical site contains a pair of spins. Let

$$A_{00} = \begin{pmatrix} 1 & 0 \\ 1 & 0 \end{pmatrix} \qquad A_{01} = \begin{pmatrix} 0 & 1 \\ 0 & 1 \end{pmatrix} \qquad A_{10} = \begin{pmatrix} 1 & 0 \\ -1 & 0 \end{pmatrix} \qquad A_{11} = \begin{pmatrix} 0 & -1 \\ 0 & 1 \end{pmatrix}, \tag{3.31}$$

or equivalently the map from virtual to physical spin-1/2 particles

$$\mathcal{P} = \begin{pmatrix} 1 & 0 & 1 & 0 \\ 0 & 1 & 0 & 1 \\ 1 & 0 & -1 & 0 \\ 0 & -1 & 0 & 1 \end{pmatrix}, \tag{3.32}$$

where the entangled pairs are in the Bell state $|\phi\rangle = |00\rangle + |11\rangle$. The map $\mathcal{P}$ corresponds to the circuit

$$\text{(3.33)}$$

Notice in this case our PEPS tensor $\mathcal{P}$ simply corresponds to unitary circuit. As such this is one of the exceptional cases in which the PEPS description *can* be considered a scalable preparation procedure.

Given an explicit MPS description of this state, we can now back out a Hamiltonian for which it is a ground state, allowing us to infer certain properties.

The initial state is constructed from entangled pairs $\prod |\phi\rangle_{2j,2j+1}$, and is the unique ground state of the Hamiltonian

$$H = -\sum_j \left( X_{2j} X_{2j+1} + Z_{2j} Z_{2j+1} \right). \tag{3.34}$$

Applying the circuit (between Bell pairs with first qubit odd and second even), we see that this transforms to

$$H' = -\sum_j \left( Z_{2j-1} X_{2j} Z_{2j+1} + Z_{2j} X_{2j+1} Z_{2j+2} \right) \tag{3.35}$$

$$= -\sum_k Z_{k-1} X_k Z_{k+1}. \tag{3.36}$$

This is precisely the cluster state Hamiltonian. The physical symmetry of this model is $\mathbb{Z}_2 \times \mathbb{Z}_2$, where $S_1 = \prod_j X_{2j-1}$ and $S_2 = \prod_j X_{2j}$. Pushing this backwards through the circuit, we see that it is equivalent to act on the virtual spins with $S_1 = \prod_j Z_{2j} Z_{2j+1}$ and $S_2 = \prod_j X_{2j} X_{2j+1}$.

This action tells us that, just like the AKLT state, the cluster state possesses SPT order.

## 3.3 MPS Properties

MPS form a vanishingly small corner of the full Hilbert space, and thus we cannot hope to use them to approximate arbitrary states. If physically relevant states correspond to those which *can* be well approximated by MPS, and MPS manifest the same non-generic properties as these physical states, then they represent an extremely useful tool to study these systems.

### 3.3.1 Decay of Correlations

We have already seen that MPS have bounded levels of entanglement, manifesting as strict area laws. We will now investigate the type of correlations which can be represented. Let $\mathcal{O}$ be some operator for which we wish to compute the two point correlator

$$\langle \psi[A] | \mathcal{O}_0 \mathcal{O}_{j+1} | \psi[A] \rangle, \tag{3.37}$$

where the subscript denotes the site at which the operator $\mathcal{O}$ is applied. Graphically this expectation value is written as:

$$\text{(3.38)}$$

We refer to the object

$$\mathbb{E}_{\mathcal{O}} = \sum_{i,j=0}^{d-1} \mathcal{O}_{i,j} A_i \otimes \bar{A}_j = \quad \text{(3.39)}$$

as the $\mathcal{O}$-transfer matrix. Note that we usually just refer to $\mathbb{E}_1$ as the transfer matrix and simply denote it $\mathbb{E}$.

The correlator (in the thermodynamic limit) can then be written as

$$\langle\psi[A]|\mathcal{O}_0\mathcal{O}_{j+1}|\psi[A]\rangle = \text{Tr}\left(\mathbb{E}^\infty\mathbb{E}_{\mathcal{O}_0}\mathbb{E}^j\mathbb{E}_{\mathcal{O}_{j+1}}\mathbb{E}^\infty\right) \tag{3.40}$$

$$\propto V_L^\dagger\mathbb{E}^j V_R. \tag{3.41}$$

where $V_L$ and $V_R$ are the dominant left and right eigenvectors of $\mathbb{E}$ respectively. The only change required when calculating longer range correlators is inserting higher powers of $\mathbb{E}$ in Eq. (3.41). The decay of correlators is therefore controlled by the eigenvalues of $\mathbb{E}$. We can normalise $A$ so that the dominant eigenvalue of $\mathbb{E}$ is 1, with the rest lying inside the unit disk. Thus any correlator can either decay exponentially with distance or be constant. Thus we see that MPS can only capture states with exponentially decaying correlations [6].

### 3.3.2 Gauge Freedom

Not all MPS represent different physical states [2]. The set of transformations of the description (i.e. the MPS) which leaves the physical state invariant are known as *gauge transformations*. In the case of MPS, these correspond to basis transformations on the virtual level:

$$|\psi[A]\rangle = \tag{3.42}$$

$$= \tag{3.43}$$

$$= |\psi[B]\rangle, \tag{3.44}$$

where $B_j = MA_jM^{-1}$. Note that $M$ is only required to have a left inverse, so can be rectangular and enlarge the bond dimension.

Another freedom is blocking. We can combine several MPS tensors $A_{i_1}, A_{i_2}, \ldots, A_{i_j}$ into a single effective tensor $B_k$, on a larger physical region

A number of canonical forms exist which partially gauge fix the MPS description. One of the most common is the left-isometric or left-canonical form (with right-isometric or right-canonical defined analogously). Here the MPS tensors obey

$$\sum_{j=0}^{d-1} A_j^\dagger A_j = \mathbb{1}_{D\times D}, \tag{3.45}$$

$$= \tag{3.46}$$

This is most useful on open boundary systems where a simple algorithm exists to put any MPS into this form. It is frequently used in numerical applications, in particular when using variational minimisation to optimise an MPS description of a ground state (DMRG), a mixed left/right isometric form is used.

Putting an MPS into this form is a *partial* gauge fixing. The remaining freedom is that of a unitary[4] on the virtual level, rather than general invertible matrix. This technique is heavily used in tensor network algorithms as a method of increasing numerical stability.

---

[4]If you include the ability to expand the bond dimension then this grows to an isometric freedom.

### 3.4 Renormalising Matrix Product States

When we renormalise a system, we usually think about attempting to write down an effective model at a longer length scale which captures the low energy portion of the original model. This can be achieved by blocking sites together, then discarding degrees of freedom to ensure the description remains useful. In the MPS, blocking can be achieved by simply contracting tensors together. How to discard only high energy degrees of freedom is a challenging question. MPS allows us to avoid having to answer this question all together [10].

Since we care only about expectation values of operators, we can work entirely in the transfer matrix picture. Blocking sites together simply consists of taking products of transfer matrices

$$\tilde{\mathbb{E}} = \mathbb{E}\mathbb{E}\mathbb{E}\mathbb{E}\mathbb{E}\ldots\mathbb{E}, \tag{3.47}$$

with sandwiched operators $\mathbb{E}_O$ being renormalised similarly. Note that the dimension of $\tilde{\mathbb{E}}$ remains $D^4$ at all times, so we never need to worry about discarding degrees of freedom. We can also use transfer matrices formed from different MPS to get off-diagonal terms of the form $\langle\psi|O|\phi\rangle$.

### 3.5 Mixed States and Many Body Operators

As described above, an MPS can be used to represent a pure state. How is a mixed state represented in this language?

Let $|\psi[A]\rangle$ be some (pure) MPS state. We can write the density matrix corresponding to $|\psi[A]\rangle$ as

$$\rho[A] = |\psi[A]\rangle\langle\psi[A]| \tag{3.48}$$

$$= \cdots \qquad\qquad\qquad\qquad\qquad \cdots . \tag{3.49}$$

The reduced density matrix on some subset of spins $R$ will therefore be represented as

$$\rho[A]_R = |\psi[A]\rangle\langle\psi[A]| \tag{3.50}$$

$$= \qquad\qquad\qquad\qquad , \tag{3.51}$$

where we have used the left and right normal forms to bring in the boundary terms.

The above network is an example of what is referred to as *matrix product operators* (MPOs) [5, 11, 12]. The general form of MPOs we will be considering is

$$\tag{3.52}$$

In addition to being used to represent density matrices, MPOs can be used to represent a large class of many body operators, including small depth quantum circuits and local Hamiltonians. For example, the transverse field Ising Hamiltonian

$$H = -J\sum X_j X_{j+1} - h\sum Z_j \tag{3.53}$$

can be represented on a line with the (operator valued) matrix

$$M = \begin{pmatrix} \mathbb{1} & 0 & 0 \\ X & 0 & 0 \\ -hZ & -JX & \mathbb{1} \end{pmatrix} \tag{3.54}$$

and end vectors

$$v_L = \begin{pmatrix} 0 & 0 & 1 \end{pmatrix} \qquad \text{and} \qquad v_R = \begin{pmatrix} 1 \\ 0 \\ 0 \end{pmatrix}. \tag{3.55}$$

The Hamiltonian on $N$ sites is then obtained as

$$H = v_L M^N v_R. \tag{3.56}$$

The Heisenberg model

$$H = -J_X \sum X_j X_{j+1} - J_Y \sum Y_j Y_{j+1} - J_Z \sum Z_j Z_{j+1} - h \sum Z_j \tag{3.57}$$

can be obtained in the same fashion with

$$v_L = \begin{pmatrix} 0 & 0 & 0 & 0 & 1 \end{pmatrix}, \qquad M = \begin{pmatrix} \mathbb{1} & 0 & 0 & 0 & 0 \\ X & 0 & 0 & 0 & 0 \\ Y & 0 & 0 & 0 & 0 \\ Z & 0 & 0 & 0 & 0 \\ -hZ & -J_X X & -J_Y Y & -J_Z Z & \mathbb{1} \end{pmatrix}, \qquad v_R = \begin{pmatrix} 1 \\ 0 \\ 0 \\ 0 \\ 0 \end{pmatrix}. \tag{3.58}$$

More generally, an MPO can be used to represent any operator which does not increase the Schmidt rank of any state too much. An existing explicit analytic construction of MPOs for 1D local Hamiltonians, as well as a new generalisation for higher dimensional Hamiltonians, is covered in more detail in Appendix A.

---

**Problems 3**

Solutions in accompanying document.

1. Describe the state given by an MPS with tensor

$$A = \begin{array}{c} \\ 00 \\ 10 \\ 01 \\ 11 \end{array} \begin{pmatrix} 0 & 1 \\ 1 & 0 \\ 0 & 1 \\ 1/2 & -1/2 \\ 1/2 & -1/2 \end{pmatrix} \qquad 1 - \boxed{A} - 3 \\ \qquad\qquad\qquad | \\ \qquad\qquad\qquad 2 \qquad , \tag{3.59}$$

where index ordering is as shown and indices 1 and 2 are combined. Boundary conditions require inserting a Pauli Z before closing periodic BCs, similar to Eq. (3.19).

2. Describe the state given by the MPS whose only nonzero components are

$$0 - \boxed{A} - 0 \;=\; 1 - \boxed{A} - 1 \;=\; 0 - \boxed{A} - 1 \;=\; 1 - \boxed{A} - 0 \;=\; 1, \tag{3.60}$$

where the left and right boundary conditions are $|0\rangle$.
*Hint: Writing out the matrices corresponding to fixing the physical index might help!*

3. Describe the qu*d*it state given by the MPS

$$i - \boxed{A} - i \oplus j \\ \qquad | \\ \qquad j \qquad\qquad = 1 \tag{3.61}$$

where $i, j \in \mathbb{Z}_d$, $\oplus$ denotes addition mod $d$, the left boundary condition is $|0\rangle$, and the right boundary is $|q\rangle$ for some $q \in \mathbb{Z}_d$.

4. Let $\mathcal{G}$ be some group. Describe the operator given by the MPO with

$$g -\boxed{M}- g \cdot h \quad = 1 \tag{3.62}$$

where the left boundary condition is $|1\rangle$, the right boundary is $|q\rangle$ for some $q \in \mathcal{G}$, and $g \cdot h$ denotes group multiplication.

5. Suppose the local basis is labelled by particle number. What is the action of the following operator (bond dimension linearly increasing left to right)?

$$n -\boxed{M}- n + m \quad = 1 \tag{3.63}$$

with left vector $L = |0\rangle$ and right vector $R = \sum_{i=0}^{N} i|i\rangle$.

6. Write an MPO for the transverse-field-cluster Hamiltonian

$$H = -J \sum_j Z_{j-1} X_j Z_{j+1} - h \sum_j X_j. \tag{3.64}$$

*Hint: This can be done with bond dimension 4.*

7. Use the ideas of MPSs and MPOs to prove that log depth quantum circuits can be simulated efficiently on a classical computer.

## 3.7 References

[1] F. Verstraete and J. Cirac, "Matrix product states represent ground states faithfully," *Physical Review B* **73**, 94423, arXiv:cond-mat/0505140v6, (2006).

[2] D. Perez-Garcia, F. Verstraete, M. M. Wolf, and J. I. Cirac, "Matrix Product State Representations," *Quantum Information & Computation* **7**, 401–430, arXiv:quant-ph/0608197, (2007).

[3] M. B. Hastings, "An area law for one-dimensional quantum systems," *Journal of Statistical Mechanics* **2007**, P08024, arXiv:0705.2024, (2007).

[4] X. Chen, Z. C. Gu, and X. G. Wen, "Local unitary transformation, long-range quantum entanglement, wave function renormalization, and topological order," *Physical Review B* **82**, 155138, arXiv:1004.3835, (2010).

[5] U. Schollwöck, "The density-matrix renormalization group in the age of matrix product states," *Annals of Physics* **326**, 96–192, arXiv:1008.3477, (2011).

[6] R. Orús, "A practical introduction to tensor networks: Matrix product states and projected entangled pair states," *Annals of Physics* **349**, 117–158, arXiv:1306.2164, (2014).

[7] F. Verstraete, V. Murg, and J. Cirac, "Matrix product states, projected entangled pair states, and variational renormalization group methods for quantum spin systems," *Advances in Physics* **57**, 143–224, arXiv:0907.2796, (2008).

[8] N. Schuch, D. Pérez-García, and I. Cirac, "Classifying quantum phases using matrix product states and projected entangled pair states," *Physical Review B* **84**, 165139, arXiv:1010.3732, (2011).

[9] I. Affleck, T. Kennedy, E. H. Lieb, and H. Tasaki, "Rigorous results on valence-bond ground states in antiferromagnets," *Physical Review Letters* **59**, 799, (1987).

[10] F. Verstraete, J. Cirac, J. Latorre, E. Rico, and M. Wolf, "Renormalization-Group Transformations on Quantum States," *Physical Review Letters* **94**, 140601, arXiv:quant-ph/0410227, (2005).

[11] B. Pirvu, V. Murg, J. I. Cirac, and F. Verstraete, "Matrix product operator representations," *New Journal of Physics* **12**, 025012, arXiv:0804.3976, (2010).

[12] I. P. McCulloch, "From density-matrix renormalization group to matrix product states," *Journal of Statistical Mechanics: Theory and Experiment* **2007**, P10014, arXiv:cond-mat/0701428, (2007).

# 4 Classifying Gapped Phases in 1D

Matrix product states are extremely useful in both analytic and numerical applications. One of the most powerful results in the field of tensor network analytics is a complete classification of gapped phases in 1D.

To begin this lecture, we will introduce quantum phases. We will then argue that in the absence of symmetry constraints, all MPS are in the same phase. Finally, we will show how symmetries change this classification. Whilst interesting in it's own right, this material also serves to demonstrate the analytic power of TNN.

## 4.1 Quantum Phases

The classical definition of a phase, or more particularly a phase transition, is usually associated to some nonanalytic behaviour of the free energy density

$$f(\beta, \mathbf{v}) = -\frac{\log \operatorname{tr} e^{-\beta H(\mathbf{v})}}{\beta}, \tag{4.1}$$

where $\mathbf{v}$ is some vector of parameters of the model (pressures, masses, coupling strengths, etc.) and $H$ the Hamiltonian of our system. Clearly when we take the quantum limit ($\beta \to \infty$), the free energy is simply the ground state energy. A *quantum phase transition* is thus associated with the ground state [1].

At a classical phase transition, correlations become long ranged

$$\langle \mathcal{O}_0 \mathcal{O}_x \rangle - \langle \mathcal{O}_0 \rangle \langle \mathcal{O}_x \rangle \sim |x|^{-\nu}, \tag{4.2}$$

where the averages are taken with respect to some thermal distribution. We therefore say that a thermal (classical) phase transition is driven by thermal fluctuations, where the variance measures the increasingly long range of these fluctuations. A quantum phase transition also has divergent correlation length, however there is no thermal average — the statistics are purely quantum in origin [1].

A classical phase corresponds to a range of deformations of $H$ and $\beta$ which can be made without causing nonanalyticities in the free energy $f$. Likewise, a quantum phase transition occurs where the ground state energy becomes nonanalytic (in the thermodynamic limit) as a function of some Hamiltonian parameters (not temperature this time!). Suppose we have a continuous family of quantum Hamiltonians $H(\lambda)$. The lowest energy levels generically act in one of the following ways [1]:

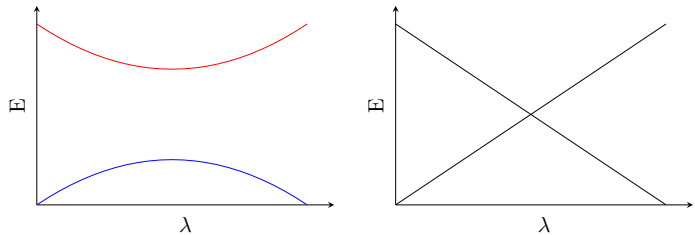

On the left, there is no phase transition, whilst on the right a transition occurs when the roles of the ground and first excited states cross.

For our purposes, a phase transition will be associated with a gapless point in the spectrum. Therefore, we will say that two states $|\psi_0\rangle$ and $|\psi_1\rangle$ are in the same phase if there is a continuous family of Hamiltonians $H(\lambda)$ such that $|\psi_0\rangle$ is the ground state of $H(0)$, $|\psi_1\rangle$ is the ground state of $H(1)$, and the gap remains open for all $\lambda \in [0, 1]$.

An equivalent notion is finite time evolution under a local Hamiltonian [2]. Two states are in the same phase if they can be interconverted by time evolution for a finite period. This is linked to the possibility of one state naturally evolving into the other.

It is simpler, and essentially equivalent, to ask which states can be interconverted by a local quantum circuit of depth constant in the system size [3, 4]. We will work within this framework. One may also ask the more complicated question of how phases change if we impose a symmetry; if we insist that all of the Hamiltonians $H(\lambda)$ commute with some symmetry group $U_g(\lambda)$. In the circuit picture, this corresponds to restricting the gate set to only gates which commute with this symmetry [4–6].

## 4.2 Injective MPS

In this lecture, we will restrict ourselves to the case of *injective* MPS [7,8]. If we assume the MPS is in left canonical form

$$\sum_{j=0}^{d-1} A_j^\dagger A_j = \mathbb{1}_{D \times D} \qquad \text{or} \qquad \text{[diagram]} = \text{[diagram]} \;, \tag{4.3}$$

then injective MPS are those for which the identity is the *unique* eigenvalue 1 left eigenvector of the transfer matrix. Moreover this means that there exists a unique full-rank[5] density matrix $\rho$ which is a 1 right eigenvector, i.e.

$$\sum_{j=0}^{d-1} A_j \rho A_j^\dagger =: \mathcal{E}(\rho) = \rho \tag{4.4}$$

$$\text{[diagram]} = \text{[diagram]} \;. \tag{4.5}$$

These MPS correspond to unique gapped ground states of local Hamiltonians [9]. The arguments we will present here generalise to non-injective MPS, however they become very technical.

## 4.3 No Topological Order

We will refer to states which cannot be connected by any constant depth local circuit as being in distinct *topological phases*, or having distinct *topological order*. This is to distinguish them from the symmetric phases we will discuss later in this lecture. In fact, we will see that there are no nontrivial topological phases in 1D [3].

Let $A_j$ define some injective MPS, and construct the transfer matrix $\mathbb{E}$[6]

$$\mathbb{E} = \text{[diagram]} \;. \tag{4.6}$$

As discussed in the previous lecture, this can be used to renormalise the MPS. Taking products of this transfer matrix corresponds to blocking sites of the original MPS. Since the MPS is injective, the leading eigenvalue of $\mathbb{E}$ is 1 and all other eigenvalues are strictly smaller. Therefore, by taking the $k$th power of the transfer matrix, we obtain a new transfer matrix which is

$$\mathbb{E}^k = \text{[diagram]} + \tilde{\mathcal{O}}\left(|\lambda_2|^k\right), \tag{4.7}$$

where $|\lambda_2| < 1$ is the second eigenvalue of the transfer matrix and $\rho$ is the fixed point of the channel. This transfer matrix can be decomposed to give a new effective MPS tensor describing the long wavelength physics

$$\tilde{A} = \text{[diagram]} \;. \tag{4.8}$$

On the regions we blocked together, we could have first applied a unitary to the state without changing the blocked transfer matrix. Since we only required a constant number of sites to be blocked to achieve this MPS tensor, this unitary freedom is restricted to a constant depth unitary circuit – precisely the equivalence we wish to allow. Now, let $V$ be some unitary which acts as $\sum_{j,k} \sqrt{\rho}_{j,k} |j, k\rangle \rightarrow$

---

[5]Were $\rho$ not full rank we could reduce the bond dimension such that it were without changing any observables in the thermodynamic limit.

[6]Note that $\mathbb{E}$ is the 'Liouville superoperator' form of the channel $\mathcal{E}$ (Eqn. 4.4)

$|0, 0\rangle$ on the state given by $\sqrt{\rho}$ and arbitrarily on the rest of the space. We can now use this to apply two circuit layers to the MPS

$$(4.9)$$

which completely disentangles the MPS, giving the state $|00 \cdots 0\rangle$.

Notice that this was all achieved by simply blocking a constant number of sites together, so we have only used a constant depth quantum circuit. Therefore, all injective MPS are in the same (topological) phase as the product state, and therefore each other.

## 4.4 Symmetry Respecting Phases

*The proofs in this section are translated into TNN from Ref. [8].*

Since there are no nontrivial topological phases, we will now examine what happens when a symmetry restriction is imposed on the allowed gates. Let $\mathcal{G}$ be some symmetry group for a state which acts on-site as $U_g := u_g^{\otimes n}$ for each $g \in \mathcal{G}$, where $u_g$ is a unitary representation of $\mathcal{G}$ acting on a single site. Recall that for $u_g$ to be a representation, we must have

$$u_g u_h = u_{gh} \tag{4.10}$$

for all $g, h \in \mathcal{G}$.

Let $A$ be an MPS tensor such that $|\psi[A]\rangle$ is symmetric, meaning that $U_g|\psi[A]\rangle = e^{i\phi_g}|\psi[A]\rangle$ for all $g \in \mathcal{G}$. We will now examine how this symmetry is realised on the MPS tensor itself.

We require an understanding of the action of unitaries on the physical level of an MPS, and when they can be 'pushed through' to act on the virtual level. There, they won't be touched by the action of constant depth symmetric circuits on the physical legs, so any properties associated with the virtual action of the symmetry will be an invariant of the phase.

We require two lemmas.

**Lemma 1.** *Let $u$ be some unitary and $A$ an injective MPS tensor. Then the largest eigenvalue $\lambda$ of the $u$-transfer matrix*

$$\mathbb{E}_u = \text{\raisebox{-1em}{}} \tag{4.11}$$

*is contained within the unit disk.*

**Proof.** Let $v^\dagger$ (note that we are not assuming that this is unitary) be a left eigenvector of $\mathbb{E}_u$

$$(4.12)$$

We therefore get for some density matrix $\rho$

$$(4.13)$$

Once again let $\rho$ be the (unique) right eigenvector of $\mathbb{E}$ with eigenvalue 1. We can view the above expression as an inner product between two vectors

$$\lambda \quad [\text{diagram}] \quad = \quad [\text{diagram}] \quad . \tag{4.14}$$

We can now apply the Cauchy-Schwarz inequality across the dotted line, giving

$$\left| [\text{diagram}] \right|^2 \quad \leq \quad [\text{diagram}] \quad \times \quad [\text{diagram}] \tag{4.15}$$

$$= \quad [\text{diagram}] \quad \times \quad [\text{diagram}] \tag{4.16}$$

$$= \quad [\text{diagram}] \quad \times \quad [\text{diagram}] \tag{4.17}$$

where the vertical lines indicate absolute value. Thus we have

$$|\lambda| \quad [\text{diagram}] \quad \leq \quad [\text{diagram}] \quad , \tag{4.18}$$

and so $|\lambda| \leq 1$. $\quad\square$

**Lemma 2.** *Equality is achieved in* <span style="color:red">*Lemma 1*</span> *if and only if there exists a unitary $v$ and an angle $\theta$ such that*

$$[\text{diagram}] \quad = e^{i\theta} \quad [\text{diagram}] \quad . \tag{4.19}$$

**Proof.** First we prove the 'if' direction. Assume that Eqn. <span style="color:red">4.19</span> holds. Then

$$[\text{diagram}] \quad = e^{i\theta} \quad [\text{diagram}] \tag{4.20}$$

$$\Longrightarrow \quad [\text{diagram}] \quad = e^{i\theta} \quad [\text{diagram}] \tag{4.21}$$

$$\Longrightarrow \quad [\text{diagram}] \quad = e^{i\theta} \quad [\text{diagram}] \quad , \tag{4.22}$$

and so we have found a left eigenvector $v^\dagger$ of $\mathbb{E}_u$ with a modulus 1 eigenvalue of $\lambda = e^{i\theta}$.

Now we prove the 'only if' direction. Assume there exists a left eigenvector $v^\dagger$ with eigenvalue of modulus 1, then the Cauchy-Schwarz inequality Eqn. 4.15 must become an equality. Therefore, there is some scalar $\alpha$ such that

$$
\begin{array}{c}
\vcenter{\hbox{}}
\end{array}
= \alpha
\begin{array}{c}
\vcenter{\hbox{}}
\end{array}
. \tag{4.23}
$$

Taking the norm of each side as vectors, we have

$$
\vcenter{\hbox{}} = |\alpha|^2 \vcenter{\hbox{}} \tag{4.24}
$$

$$
\implies \vcenter{\hbox{}} = |\alpha|^2 \vcenter{\hbox{}} \tag{4.25}
$$

$$
\implies \vcenter{\hbox{}} = |\alpha|^2 \vcenter{\hbox{}} . \tag{4.26}
$$

Therefore, $|\alpha| = 1$, so $\alpha = e^{i\theta}$.

Since $\rho$ is full rank, it is invertible, so

$$
\vcenter{\hbox{}} = e^{i\theta} \vcenter{\hbox{}} . \tag{4.27}
$$

Now, rearranging this and left multiplying by $v^\dagger$, we have

$$
\vcenter{\hbox{}} = e^{i\theta} \vcenter{\hbox{}} \tag{4.28}
$$

$$
\implies \vcenter{\hbox{}} = e^{i\theta} \vcenter{\hbox{}} \tag{4.29}
$$

$$
\implies \lambda \vcenter{\hbox{}} = e^{i\theta} \vcenter{\hbox{}} . \tag{4.30}
$$

We therefore see that $v^\dagger v$ is a left eigenvector of the transfer matrix $\mathbb{E}$ with norm-1 eigenvalue. By assuming injectivity however we require that the only norm-1 eigenvalue is the non-degenerate $+1$ eigenvalue, whose left eigenvector is the identity. Thus we conclude $v$ is, after rescaling, unitary, and that Eqn. 4.19 therefore holds. $\square$

So far, we have established that a unitary $u$ can be 'pushed through' the MPS tensor if and only if the $u$-transfer matrix has an eigenvalue of unit magnitude. We will now show that $u$ is a local symmetry if and only if it can be pushed through. This will complete our understanding of the action of local symmetries on MPS tensors.

**Theorem 1** (Symmetries push through). *Let $\mathcal{G}$ be a group. A unitary representation $u_g$ is a local symmetry if and only if*

$$
\begin{array}{c}\text{\includegraphics}\end{array} = e^{i\theta_g} \begin{array}{c}\text{\includegraphics}\end{array} \tag{4.31}
$$

*for $v_g$ unitary and $\theta_g \in [0, 2\pi)$.*

**Proof.** If Eqn. 4.31 holds, it is clear that $u_g$ is a symmetry since $v_g$ is simply a gauge transformation on the MPS.

Let

$$
\sigma_k = \begin{array}{c}\text{\includegraphics}\end{array} \tag{4.32}
$$

be the reduced density matrix on $k$ sites, where $\rho$ is the right fixed point of $\mathbb{E}$. By construction, $\operatorname{tr}(\sigma_k) = 1$, but $\sigma_k$ will generically be mixed, so $\operatorname{tr}(\sigma_k^2) \leq 1$. Recall that the purity of a density matrix is lower bounded by the inverse of the matrix-rank, i.e. $\operatorname{tr}(\sigma_k^2) \geq 1/\operatorname{rank}(\sigma_k)$. Since our reduced density matrix is obtained from a bond dimension $D$ MPS, it has rank at most $D^2$. Therefore

$$
\frac{1}{D^2} \leq \operatorname{tr}\left(\sigma_k^2\right) = \begin{array}{c}\text{\includegraphics}\end{array} \tag{4.33}
$$

$$
= \begin{array}{c}\text{\includegraphics}\end{array} \tag{4.34}
$$

$$
= \begin{array}{c}\text{\includegraphics}\end{array}, \tag{4.35}
$$

where the second equality holds because $u_g$ is a local symmetry.

Here, the left and right boundary vectors ($\mathbb{1}$ and $\rho$) are independent of the number of sites upon which $\sigma_k$ is supported, so this inequality holds for all $k$. This can only be the case if $\mathbb{E}_{u_g}$ has an eigenvalue of magnitude 1, as it would otherwise have to possess exponential decay. From Lemma 2, this implies that there exists some unitary $v_g$ and an angle $\theta_g$ such that

$$\vcenter{\hbox{\includegraphics{eq436left}}} = e^{i\theta_g}\,\vcenter{\hbox{\includegraphics{eq436right}}} \tag{4.36}$$

which completes the proof. $\square$

We now investigate the properties of the virtual action of the symmetry. As discussed above, if we apply a constant depth circuit with symmetric gates to the MPS (i.e. mapping us to any other state in the phase), we can push the symmetry action first through the circuit and then onto the virtual level. Therefore, any properties it has will be an invariant of the phase.

<div style="border:1px solid #000; padding:10px;">

**Aside 2: Projective representations**

Let $\mathcal{G}$ be some group. A (linear) representation $u_g$ obeys

$$u_g u_h = u_{gh} \quad \forall g, h \in \mathcal{G}. \tag{4.37}$$

This is not the most general way of acting with a group however. We could also ask for

$$v_g v_h = \omega[g, h] v_{gh} \quad \forall g, h \in \mathcal{G}, \tag{4.38}$$

where $\omega[g, h] = e^{i\phi[g,h]}$ is a scalar which depends on both $g$ and $h$ independently. This is known as a projective representation. One might ask whether this is simply a more complicated way of writing a linear representation. Maybe we can rephase $v_g$ to obtain Eqn. 4.37. Let $\beta[g]$ be some phase depending only on $g$ then after a rephasing $v_g \mapsto \beta[g] v_g$, we have

$$v_g v_h = \omega[g, h] \frac{\beta[gh]}{\beta[g]\beta[h]} v_{gh} = \omega'[g, h] v_{gh}. \tag{4.39}$$

We say that $\omega$ and $\omega'$ are equivalent if they are related in this way, so

$$\omega \sim \omega' \iff \exists \beta : \omega'[g, h] = \frac{\beta[gh]}{\beta[g]\beta[h]} \omega[g, h]. \tag{4.40}$$

A projective representation is therefore equivalent to a linear representation if the phases can be completely removed, i.e. there exists a $\beta$ such that

$$\omega[g, h] = \frac{\beta[g]\beta[h]}{\beta[gh]}. \tag{4.41}$$

As you will show in Problems 4, there are projective representations which are *not* equivalent to any linear representation.

</div>

Suppose we act with $u_g$ followed by $u_h$ on the MPS tensor, then

$$\vcenter{\hbox{\includegraphics{eq442a}}} = e^{i\theta_g}\,\vcenter{\hbox{\includegraphics{eq442b}}} = e^{i\theta_g} e^{i\theta_h}\,\vcenter{\hbox{\includegraphics{eq442c}}}. \tag{4.42}$$

We could also have combined $u_g u_h = u_{gh}$ before pushing through, which tells us

$$\vcenter{\hbox{\includegraphics{eq443a}}} = e^{i\theta_{gh}}\,\vcenter{\hbox{\includegraphics{eq443b}}}. \tag{4.43}$$

Therefore

$$(v_g \otimes v_g^\dagger)(v_h \otimes v_h^\dagger) = \frac{e^{i\theta_{gh}}}{e^{i\theta_g}e^{i\theta_h}} v_{gh} \otimes v_{gh}^\dagger, \tag{4.44}$$

so $(v_g \otimes v_g^\dagger)$ is equivalent to a linear representation. We can split this across the tensor product, telling us that in general

$$v_g v_h = \omega[g,h] v_{gh}, \tag{4.45}$$

where $\omega$ is some phase. We cannot say anything about the phase in this case, since *anything* would be cancelled by tensoring with the conjugate.

The only freedom we have to change $v_g$ within a phase is local rephasing, therefore the equivalence classes of $\omega$ label the different phases of injective MPS with a symmetry restriction. These equivalence classes are indexed by the so-called second group cohomology class of the group $\mathcal{G}$, an object usually written as $\mathcal{H}^2(\mathcal{G}, U(1))$ [2, 10].

---

**Problems 4**

Solutions in accompanying document.

1. The group $\mathbb{Z}_2 \times \mathbb{Z}_2$ has the presentation $\mathbb{Z}_2 \times \mathbb{Z}_2 = \langle x, z | x^2 = z^2 = e, xz = zx \rangle$. Show that the Pauli matrices form a projective representation of $\mathbb{Z}_2 \times \mathbb{Z}_2$.

   *Hint: let $v_x = X$, $v_z = Z$, $v_{xz=zx} = Y$ and show that $v_g v_h = \omega[g,h] v_{gh}$, where $\omega$ is some phase.*

2. Determine the *factor system* $\omega[g,h]$ for the Pauli matrices.

3. Show that the Pauli projective representation is not equivalent to a linear representation.

   *Hint: $xz = zx$, can we rephase $v_x$ and $v_z$ to make $v_x v_z - v_z v_x = 0$?*

4. Recall from Sec. 3.2 that the symmetry of the cluster state is $\mathbb{Z}_2 \times \mathbb{Z}_2$, with the action on the MPS tensor being

   

   $$\tag{4.46}$$

   What can we conclude about the cluster state?

---

## 4.6 References

[1] S. Sachdev, Quantum phase transitions 2nd edition, Cambridge University Press, (2011).

[2] X. Chen, Z. C. Gu, and X. G. Wen, "Classification of Gapped Symmetric Phases in 1D Spin Systems," *Physical Review B* **83**, 035107, arXiv:1008.3745, (2011).

[3] X. Chen, Z. C. Gu, and X. G. Wen, "Local unitary transformation, long-range quantum entanglement, wave function renormalization, and topological order," *Physical Review B* **82**, 155138, arXiv:1004.3835, (2010).

[4] Y. Huang, and X. Chen, "Quantum circuit complexity of one-dimensional topological phases," *Physical Review B* **91**, 195143, arXiv:1401.3820, (2015).

[5] X. Chen, Z. C. Gu, Z. X. Liu, and X. G. Wen, "Symmetry protected topological orders and the group cohomology of their symmetry group," *Physical Review B* **87**, 155114, arXiv:1106.4772, (2011).

[6] "Can a symmetry-preserving unitary transformation that goes from a trivial SPT to a non-trivial SPT be local?," Stack Exchange - http://physics.stackexchange.com/questions/184570/can-a-symmetry-preserving-unitary-transformation-that-goes-from-a-trivial-spt-to, (2015).

[7] D. Perez-Garcia, F. Verstraete, M. M. Wolf, and J. I. Cirac, "Matrix Product State Representations," *Quantum Information & Computation* **7**, 401–430, arXiv:quant-ph/0608197, (2007).

[8] D. Pérez-García, M. M. Wolf, M. Sanz, F. Verstraete, and J. Cirac, "String Order and Symmetries in Quantum Spin Lattices," *Physical Review Letters* **100**, 167202, arXiv:0802.0447v1, (2008).

[9] N. Schuch, I. Cirac, and D. Pérez-García, "PEPS as ground states: Degeneracy and topology," *Annals of Physics* **325**, 2153–2192, arXiv:1001.3807, (2010).

[10] N. Schuch, D. Pérez-García, and I. Cirac, "Classifying quantum phases using matrix product states and projected entangled pair states," *Physical Review B* **84**, 165139, arXiv:1010.3732, (2011).

# 5  Tensor network algorithms

One area in which tensor networks have had exceptional practical success is in low-temperature simulation of condensed matter systems. A relatively well-understood toy model is finding ground states of one-dimensional spin systems. Even under the assumption of a local Hamiltonian, this seemingly narrow problem retains QMA-completeness [1] (a quantum analogue of NP), dashing any hope of general simulation, even on a quantum computer. Whilst this may at first seem like a significant problem, many 'physically realistic' systems don't exhibit this prohibitive complexity. Tensor networks can be used to exploit, and to a certain extent understand, this structure.

As discussed previously, states of low entanglement are well represented in the form of MPS. If we consider the case of local and *gapped* Hamiltonians, it has been shown that the relevant ground states cannot be highly entangled [2–5, 12] (see Ref. [6] for a review). This restricted entanglement means that such states admit efficient MPS approximations [7], and moreover that they may be efficiently approximated [8–12], showing that the presence of the gap causes the complexity to plummet from QMA-complete all the way down to P, removing the complexity barrier to simulation. We note that despite the challenges, both complexity theoretic and physical, in applying MPS to gapless models, they have been successfully utilised for this purpose [13–15].

More concretely, the way in which we plan to approximate the ground state is by minimising the Rayleigh quotient of the Hamiltonian $H$ (the energy) over some restricted domain $\mathcal{D}$ to yield an approximate ground state $|\Gamma\rangle$ given as

$$|\Gamma\rangle := \arg\min_{|\psi\rangle \in \mathcal{D}} \frac{\langle\psi|H|\psi\rangle}{\langle\psi|\psi\rangle}. \tag{5.1}$$

As we know that the exact solution is well-approximated by MPS, we will restrict ourselves to the domain $\mathcal{D}$ of MPS of a bounded bond dimension. The idea behind DMRG and TEBD is to start in some MPS state[7] then variationally move along this domain, minimising the energy as we go. The difference between both methods is the manner in which this variation step is performed, with DMRG and TEBD taking more computational and physical approaches respectively.

Although the algorithms we discuss here are designed for finding MPS ground states, they can be adaped to simulate time evolution [16, 17], find Gibbs states [18], or optimise other operators acting on a statespace of interest [19].

## 5.1  DMRG (The Computer Scientist's approach)

By far the most studied and successful of the algorithms in the field is DMRG. For clarity we will be restricting ourselves to finite DMRG, though there do exist thermodynamic variants. DMRG is an umbrella term which encompasses several similar algorithms, the algorithm we will discuss here is a simplified but nonetheless effective example. As the introduction of this algorithm in Ref. [20] pre-dates TNN, its description has historically been presented in a far more physically motivated and technically complicated manner. Due to the corresponding shift in interpretation, the original acronym now holds little relevance to the modern tensor network interpretation of DMRG, and so for clarity we intentionally omit defining precisely the expansion of DMRG as an acronym[8]. For a full review in pre-TNN notation see Ref. [21], and see Ref. [22] for a TNN treatment.

Representing the Hamiltonian by an MPO, optimising the Rayleigh quotient over MPS looks like the following:

$$\tag{5.2}$$

The difficulty is that as we need the contraction of these MPS tensors; the overall objective function is highly non-linear, but it does however only depend quadratically on each individual tensor. The key heuristic behind DMRG is to exploit the simplicity of these local problems, approximating the multivariate (multi-tensor) optimisation by iterated univariate (single tensor) optimisations.

---

[7]Typically a random MPS is sufficient in practice, though one could use an educated guess if available.

[8]Though a curious reader is free to Google it, at their own peril.

Note that while the DMRG algorithm we are going to outline only calculates ground states, related generalisations exist which can be used to simulate excited states, dynamics etc.

### 5.1.1 One-site

The simplest interpretation of the above sketch of DMRG is known as DMRG1 (or one-site DMRG). For a fixed site $i$, the sub-step involves fixing all but a single MPS tensor, which is in turn optimised over, i.e.

$$A_i \longleftarrow \underset{A_i}{\arg\min} \frac{\langle \psi(A_i)|H|\psi(A_i)\rangle}{\langle \psi(A_i)|\psi(A_i)\rangle}. \tag{5.3}$$

In TNN these step look like:

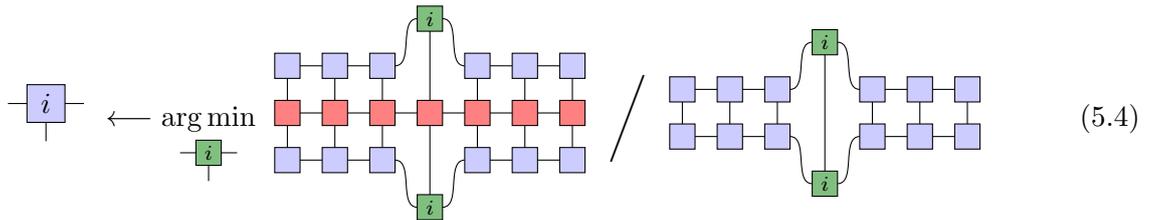

$$\tag{5.4}$$

Next we define the *environment tensors*

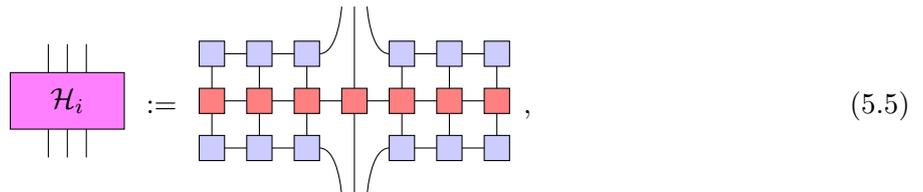

$$\tag{5.5}$$

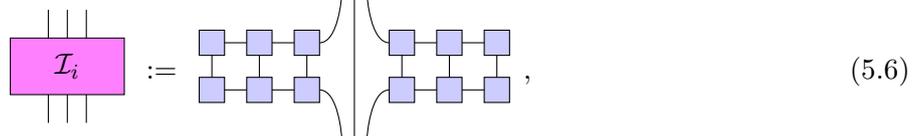

$$\tag{5.6}$$

which correspond to taking closed tensor networks — the expectation values of $H$ and the $I$ respectively — and removing the objective tensor. Given these environments, the sub-step in Eq. (5.4) becomes

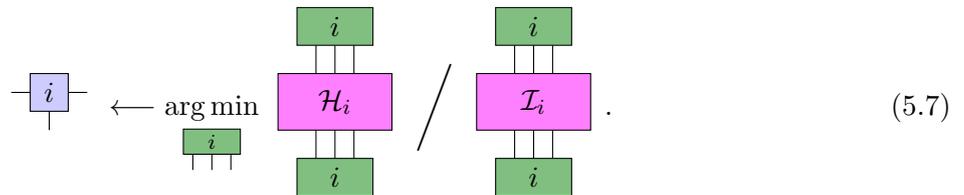

$$\tag{5.7}$$

Vectorising this equation yields

$$A_i \longleftarrow \underset{A_i}{\arg\min} \frac{\langle A_i|\mathcal{H}_i|A_i\rangle}{\langle A_i|\mathcal{I}_i|A_i\rangle}. \tag{5.8}$$

Finally we can simplify the denominator of this objective function by appropriately gauge-fixing our MPS to be in canonical form. By putting the parts of the MPS left of our site in left-canonical form, and those to the right in right-canonical form, then we get that $\mathcal{I}_i$ simply reduces to the identity:

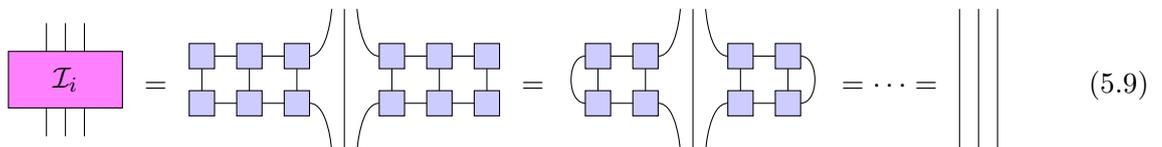

$$\tag{5.9}$$

Given this canonicalisation, the problem thus reduces to

$$A_i \longleftarrow \underset{A_i}{\arg\min} \frac{\langle A_i | \mathcal{H}_i | A_i \rangle}{\langle A_i | A_i \rangle}. \tag{5.10}$$

As $\mathcal{H}_i$ is Hermitian, this optimisation has a closed form solution given by the minimum eigenvector[9] of $\mathcal{H}_i$. By sweeping back and forth along the chain, solving this localised eigenvector problem, and then shifting along the canonicalisation as necessary, we complete our description of the algorithm.

The main advantage of DMRG1 is that the state stays within the MPS manifold without the bond dimension growing, meaning that the algorithm is greedy[10]. This strict restriction on the bond dimension can however be a double-edged sword; this means that there is no particularly convenient method of gently growing the bond dimension as the algorithm runs[11], and no information is gained regarding the appropriateness of the choice of bond dimension. Both of these problems are addressed in turn by the improved, albeit slightly more complicated, DMRG2 algorithm.

### 5.1.2 Two-site

The idea with DMRG2 is to block two sites together, perform an optimisation in the vein DMRG1, then split the sites back out. This splitting process gives DMRG2 its power, allowing for dynamic control of the bond dimension, as well as providing information about the amount of error caused by trimming, which helps to inform the choice of bond-dimension.

First an optimisation is performed:

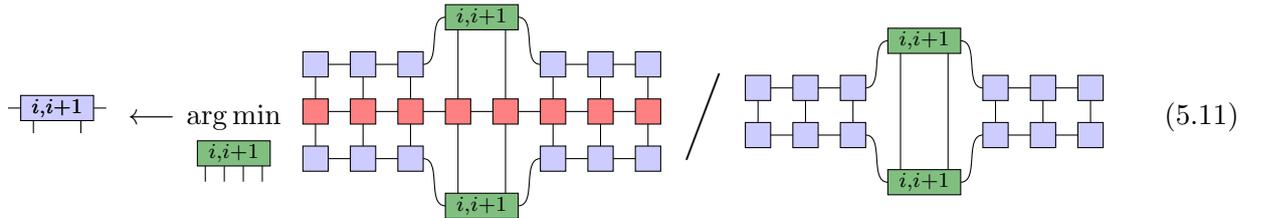

which can once again be solved by taking the minimum eigenvector of an environment tensor with respect to two sites, $\mathcal{H}_{i,i+1}$, once again in mixed canonical form. After this the two-site tensor is split apart by performing an SVD[12] and a bond trimming:

This trimmed SVD has two key features. Firstly the bond dimension to which we trim could be higher than that we originally started with, allowing us to gently expand out into the space of higher bond dimension MPS. Secondly we can use the truncated singular values to quantify the error associated with this projection back down into the lower bond dimension space, better informing our choice of bond dimension.

## 5.2 TEBD (The Physicist's approach)

Time-evolving block decimation (TEBD) [27, 28] is a tensor network algorithm that allows the dynamics of 1D spin systems to be simulated. By simulating *imaginary*-time-evolution low-temperature features such as the ground state may be calculated as well.

---

[9]If we had not canonicalised the MPS then a closed form solution still exists in the form of the *generalised* eigenvector of $\mathcal{H}_i$ and $\mathcal{I}_i$, but in general the cost of canonicalisation is well-justified by the increased stability it yields.

[10]A greedy algorithm is one which solves local problems, such that the cost function (energy in this case) monotonically decreases.

[11]There are however somewhat involved methods that allow for auxiliary data to be injected in a non-local fashion such as Refs. [23, 24] (see Ref. [25] for a review), achieving a similar goal.

[12]Whilst other factorisations such as QR and LU can also be used, SVD is preferred over other rank-revealing decompositions due to the optimality of singular value truncation as a low-rank approximation (see Aside 1).

To simulate imaginary-time-evolution, we need to approximate the imaginary-time-evolution operator $U(\tau) = \exp(-\tau H)$. The problem here is that whilst we may have an efficient representation of $H$, any exponential of it will not necessarily have a succinct representation. Take the example of a two-body Hamiltonian with corresponding imaginary-time-evolution operator

$$U(\tau) = e^{-\tau \sum_i h_i} \qquad \text{where} \qquad H = \sum_i h_i$$

and $h_i$ is an interaction term acting on spins $i$ and $i + 1$. Whilst $H$ has a constant Schmidt rank, admitting an efficient representation as an MPO, $U(\tau)$ generically has exponential bond dimension for almost all $\tau$.

Let $H_o(H_e)$ denote the sum of terms $h_i$ for odd(even) $i$. As all the terms within $H_o(H_e)$ are commuting, $e^{-\tau H_o}(e^{-\tau H_e})$ can be efficiently computed and represented. The problem of approximating $U(\tau)$ can therefore be reduced to the problem of approximating $e^{-t(A+B)}$ when only terms of the form $e^{-\tau A}$ and $e^{-\tau B}$ can be computed.

The central mathematical tool to TEBD are the *exponential product approximations*. The first order of these approximation is the Suzuki-Trotter formula, which approximates the total evolution by simply evolving each subsystem

$$e^{-\tau(A+B)} = e^{-\tau A} e^{-\tau B} + \mathcal{O}(\tau^2).$$

It turns out there exist entire families of such approximations [29], though for our purposes we will just illustrate the procedure for Suzuki-Trotter.

The TEBD algorithm works by approximating the imaginary-time-evolution operator by the above exponential product formulae, applying it to a given MPS, and trimming the bond dimension to project back down into the space of MPS.

Our approximation to the imaginary-time-evolution operator is given by a product of layers containing only nearest-neighbour two-site operators, meaning we need only be able to contract these operators into our MPS. Suppose we want to apply an operator $U$ to the spins at sites $i$ and $i + 1$. The idea is to apply the operator, contract everything into a single tensor, then once again use an SVD trimming to truncate the bond dimension back down.

(5.12)

The benefits this trimming procedure gave to DMRG2 — namely control over bond dimension growth and quantification of trimming errors — are also seen in TEBD. As the above procedure is entirely localised, TEBD also admits a large amount of parallelisation, not typically available to DMRG.

## 5.3 Implementation

From-scratch implementation of these simple algorithms can be achieved with relative ease, however several high performance libraries exist for research level simulations. We direct the interested reader to investigate ITensor [30] (C++), evoMPS [31] (Python), Matrix Product Toolkit [32] (C++), uni10 (C++) [33], Tensor Operations [34] (Julia) among others. A simple tensor class can also be easily written in MATLAB.

### Problems 5

Solutions in accompanying document.

1. Consider the critical transverse Ising model

$$H = -\sum_{i=1}^{n-1} X_i X_{i+1} - \sum_{i=1}^{n} Z_i. \tag{5.13}$$

For open boundary conditions, it is known that the ground state energy as a function of $n$ has the form [35]

$$E(n) = 1 - \csc\left(\frac{\pi}{\alpha n + \beta}\right) \tag{5.14}$$

for some integers $\alpha$ and $\beta$. Using either DMRG or TEBD, estimate the ground state energy for several chain lengths and calculate $\alpha$ and $\beta$.

2. It is known that the LOCAL HAMILTONIAN problem is in P for 1D gapped Hamiltonians [8–12]. DMRG and TEBD are the most common techniques for numerically finding the ground states of such systems. For a gapped and 1D local Hamiltonian, prove that DMRG or TEBD converge.

## 5.5 References

[1] A. Y. Kitaev, A. H. Shen, and M. N. Vyalyi, *Classical and Quantum Computation*, Graduate Studies in Mathematics, AMS, (2002).

[2] M. B. Hastings, "An area law for one-dimensional quantum systems," *Journal of Statistical Mechanics* **2007**, P08024, arXiv:0705.2024, (2007).

[3] I. Arad, Z. Landau, and U. Vazirani, "Improved one-dimensional area law for frustration-free systems," *Physical Review B* **85**, 195145, arXiv:1111.2970, (2012).

[4] I. Arad, A. Kitaev, Z. Landau, and U. Vazirani, "An area law and sub-exponential algorithm for 1D systems," arXiv:1301.1162, (2013).

[5] Y. Huang, "Area law in one dimension: Renyi entropy and degenerate ground states," arXiv:1403.0327, (2014).

[6] J. Eisert, M. Cramer, and M. B. Plenio, "Colloquium: Area laws for the entanglement entropy," *Reviews of Modern Physics* **82**, 277–306, arXiv:0808.3773, (2010).

[7] F. Verstraete and J. I. Cirac, "Matrix product states represent ground states faithfully," *Physical Review B* **73**, 094423, arXiv:cond-mat/0505140, (2006).

[8] Z. Landau, U. Vazirani, and T. Vidick, "A polynomial-time algorithm for the ground state of 1D gapped local Hamiltonians," *Proceedings of the 5th Conference on Innovations in Theoretical Computer Science* and *Nature Physics* **11**, 566–569, arXiv:1307.5143, (2013).

[9] C. T. Chubb and S. T. Flammia, "Computing the Degenerate Ground Space of Gapped Spin Chains in Polynomial Time," *Chicago Journal of Theoretical Computer Science* **2016**, 9, arXiv:1502.06967, (2016).

[10] Y. Huang, "A polynomial-time algorithm for approximating the ground state of 1D gapped Hamiltonians," arXiv:1406.6355, (2014).

[11] Y. Huang, "Computing energy density in one dimension," arXiv:1505.00772, (2015).

[12] I. Arad, Z. Landau, U. Vazirani, and T. Vidick, "Rigorous RG algorithms and area laws for low energy eigenstates in 1D," arXiv:1602.08828, (2016).

[13] L. Tagliacozzo, T. R. de Oliveira, S. Iblisdir, and J. I. Latorre, "Scaling of entanglement support for Matrix Product States," *Physical Review B* **78**, 024410, arXiv:0712.1976, (2008).

[14] F. Pollmann, S. Mukerjee, A. Turner, and J. E. Moore, "Theory of finite-entanglement scaling at one-dimensional quantum critical points," *Physical Review Letters* **102**, 255701, arXiv:0812.2903, (2008).

[15] V. Stojevic, J. Haegeman, I. P. McCulloch, L. Tagliacozzo, and F. Verstraete, "Conformal Data from Finite Entanglement Scaling," *Physical Review B* **91**, 035120, arXiv:1401.7654, (2015).

[16] A. J. Daley, C. Kollath, U. Schollwoeck, and G. Vidal, "Time-dependent density-matrix renormalization-group using adaptive effective Hilbert spaces," *Journal of Statistical Mechanics: Theory and Experiment* **2004**, P04005, arXiv:cond-mat/0403313, (2004).

[17] S. R. White, and A. E. Feiguin, "Real-Time Evolution Using the Density Matrix Renormalization Group," *Physical Review Letters* **93**, 076401, arXiv:cond-mat/0403310, (2004).

[18] F. Verstraete, J. J. García-Ripoll, and J. I. Cirac, "Matrix Product Density Operators: Simulation of finite-T and dissipative systems," *Physical Review Letters* **93**, 207204, arXiv:cond-mat/0406426, (2004).

[19] J. C. Bridgeman, S. T. Flammia, and D. Poulin, "Detecting Topological Order with Ribbon Operators," *Physical Review B* **93**, 207204, arXiv:1603.02275, (2016).

[20] S. R. White, "Density matrix formulation for quantum renormalization groups," *Physical Review Letters* **69**, 2863–2866, (1992).

[21] U. Schollwöck, "The density-matrix renormalization group," *Reviews of Modern Physics* **77**, 259–315, arXiv:cond-mat/0409292, (2005).

[22] U. Schollwöck, "The density-matrix renormalization group in the age of matrix product states," *Annals of Physics* **326**, 96–192, arXiv:1008.3477, (2011).

[23] S. R. White, "Density matrix renormalization group algorithms with a single center site," *Physical Review B* **72**, 180403, arXiv:cond-mat/0508709, (2005).

[24] S. V. Dolgov and D. V. Savostyanov, "Alternating Minimal Energy Methods for Linear Systems in Higher Dimensions," *SIAM Journal on Scientific Computing* **36**, A2248–A2271, arXiv:1301.6068, (2014).

[25] S. V. Dolgov and D. V. Savostyanov, "One-site density matrix renormalization group and alternating minimum energy algorithm," *Numerical Mathematics and Advanced Applications*, arXiv:1312.6542, (2013).

[26] C. Eckart and G. Young, "The approximation of one matrix by another of lower rank," *Psychometrika* **1**, (1936).

[27] G. Vidal, "Efficient Classical Simulation of Slightly Entangled Quantum Computations," *Physical Review Letters* **91**, 147902, arXiv:quant-ph/0301063, (2003).

[28] G. Vidal, "Efficient Simulation of One-Dimensional Quantum Many-Body Systems," *Physical Review Letters* **93**, 40502, arXiv:quant-ph/0310089, (2004).

[29] N. Hatano and M. Suzuki, "Finding Exponential Product Formulas of Higher Orders," *Quantum Annealing and Other Optimization Methods*, arXiv:math-ph/0506007, (2005).

[30] ITensor *http://itensor.org/*

[31] evoMPS *http://amilsted.github.io/evoMPS/*

[32] Matrix Product Toolkit *http://physics.uq.edu.au/people/ianmcc/mptoolkit/index.php*

[33] uni10 *http://uni10.org/*

[34] Tensor Operations *https://github.com/Jutho/TensorOperations.jl*

[35] N. S. Izmailian and C.-K. Hu, *"Boundary conditions and amplitude ratios for finite-size corrections of a one-dimensional quantum spin model,"* Nuclear Physics B **808**, 613, arXiv:1005.1710, (2009).

# 6 Projected Entangled Pair States

Many of the ideas behind MPS generalise to higher dimensions via *projected entangled pair states* or *PEPS* [1,2]. We will see how this is a misnomer in two ways, there is not necessarily a projector and there is not necessarily an entangled pair.

We begin by recalling the PEPS description of matrix product states, then generalise this to two dimensional models. After giving several examples, we will examine the properties of PEPS, identifying both the similarities and differences to MPS.

## 6.1 One Dimensional Systems: MPS

We have already seen the PEPS construction in 1D. Let $|\phi\rangle \in \mathbb{C}^D \otimes \mathbb{C}^D$ be some (usually) entangled pair and $\mathcal{P}: \mathbb{C}^D \otimes \mathbb{C}^D \to \mathbb{C}^d$ some linear map. Then

$$|\psi\rangle = \qquad\qquad\qquad\qquad\qquad\qquad\qquad\qquad\qquad\qquad\qquad\qquad \tag{6.1}$$

where

$$|\phi\rangle = \quad\bullet\!\!\sim\!\!\sim\!\!\bullet \tag{6.2}$$

is the chosen entangled pair. As we saw, we have a large choice in the exact description we use. We can transform the local basis of each spin in the entangled pair by any (left) invertible matrix

$$|\phi\rangle \to (A \otimes B)|\phi\rangle, \tag{6.3}$$

since we can modify $\mathcal{P}$ to compensate

$$\mathcal{P} \to \mathcal{P}(B^{-1} \otimes A^{-1}). \tag{6.4}$$

One thing to note is that $|\phi\rangle$ does not necessarily need to be a valid quantum state. We usually leave it unnormalised for convenience.

In addition to this gauge freedom, we have additional choices in the description. We could use entangled triplets for example. Let $|\psi\rangle = |000\rangle + |111\rangle$, then we could choose our PEPS to be

$$|\psi\rangle = \qquad\qquad\qquad\qquad\qquad\qquad\qquad\qquad\qquad\qquad\qquad\qquad \tag{6.5}$$

Clearly this doesn't offer any more descriptive power than using entangled pairs. Suppose we have some PEPS projector $\mathcal{Q}$ acting on pairs, then we can extend this to a $\mathcal{P}$ acting on triplets by

$$\mathcal{P} = \mathcal{Q}\left(\mathbb{1} \otimes (|0\rangle\langle 00| + |1\rangle\langle 11|)\right). \tag{6.6}$$

In the other direction, we can build a product of triplets using a minor modification of the GHZ MPS presented above and then use $\mathcal{Q}$ to build our state of interest.

## 6.2 Extending to Higher Dimensions

The extension from one to higher dimensional systems proceeds straightforwardly. We will discuss the simple case of a hypercubic lattice, but the framework can be carried out on any graph. In particular, we will restrict to 2D.

As before, we allow $|\phi\rangle$ to be some entangled pair. The PEPS is built as the natural generalisation to 2D

$$|\psi\rangle = \tag{6.7}$$

where

$$\mathcal{P} : \left(\mathbb{C}^D\right)^{\otimes 4} \to \mathbb{C}^d \tag{6.8}$$

is some linear operator from the virtual to the physical space.

Clearly there is a large amount of gauge freedom in this description as there was in the 1D case. Any invertible transformation of each virtual spin can be compensated in the definition of the PEPS 'projector' $\mathcal{P}$, analogous to Eq. (6.4).

As in the MPS, one may ask whether using different entanglement structures leads to greater descriptive power. It is easy to see that this is not the case in general. Suppose we choose to lay down plaquettes in a GHZ state and then act with PEPS projectors between plaquettes.

$$|\psi\rangle = \qquad , \quad \text{where} \quad \begin{array}{c}\square\end{array} = \sum_i \left|\begin{array}{cc} i & i \\ i & i \end{array}\right\rangle . \tag{6.9}$$

We can use a standard PEPS to prepare this resource state, so any state which can be prepared from this 'projected entangled plaquette' construction can be prepared from a PEPS at small additional cost.

## 6.3 Some PEPS examples

We will now look at several example PEPs.

**Product State**

We have already seen this example in 1D. Exactly the same thing works in 2D, for example take

$$\mathcal{P} = |0\rangle \left\langle 0 \quad \begin{array}{cc} 0 & \\ & 0 \end{array} \quad 0\right| . \tag{6.10}$$

**GHZ State**

Directly generalising the 1D case, we can use

$$\mathcal{P} = |0\rangle \left\langle 0 \begin{array}{cc} & 0 \\ 0 & \end{array} 0 \right| + |1\rangle \left\langle 1 \begin{array}{cc} & 1 \\ 1 & \end{array} 1 \right| \tag{6.11}$$

to build the GHZ state.

**RVB State**

Let $D = 3$ be the bond dimension and let

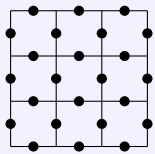

$$= 1 \tag{6.12}$$

for $\alpha \in \{1, 2\}$, as well as all rotations on the virtual level, be the only nonzero elements of the PEPS tensor. Suppose we tile these tensors and project the dangling indices onto the $|2\rangle$ state. What is the resulting physical state?

This state is known as the resonating valence bond state [2–4] and consists of a superposition of all complete tilings of the lattice with maximally entangled pairs

$$\left| \begin{array}{cccc} | & | & | & | \\ | & | & | & | \end{array} \right\rangle + \left| \begin{array}{cccc} | & | & | & | \\ = & = & | & | \end{array} \right\rangle + \left| \begin{array}{cccc} | & = & | \\ = & | & | \end{array} \right\rangle + \left| \begin{array}{cccc} = & = \\ = & = \end{array} \right\rangle + \cdots ,$$

where

$$\bullet\!\!-\!\!\bullet = |00\rangle + |11\rangle.$$

---

**Aside 3: Kitaev's Toric code**

Kitaev's Toric code [5] is a canonical example of a topologically ordered model Here we will construct a Hamiltonian with the code space as the ground space of the model. The ground state of this Hamiltonian is the superposition of all closed loops of flipped spins.

We place qubits on the edges of a square lattice.

We wish to create a Hamiltonian with closed loop states (of flipped spins) as the ground state. Suppose all spins are initially in the $|0\rangle$ state. Then around every vertex $v$ place an interaction

$$A_v = - \begin{array}{c} Z \\ -Z \!\!+\!\! Z- \\ Z \end{array} . \tag{6.13}$$

To be in the ground state of this term, the number of edges flipped to $|1\rangle$ neighbouring a given vertex must be even. Drawing edges carrying flipped spins in red, we can trace the effect of this

on the lattice

.

We can see that on a square graph, requiring an even number of edges incident on each vertex enforces that all of our loops are closed.

At this point, our ground space contains all states with only closed loops. We want an equal superposition of all closed loop states. This is achieved by placing an interaction around plaquettes or squares on the lattice, which convert between loop states. To be an eigenstate, all loop states reachable from the vacuum state must be in the superposition. At each plaquette $p$, place an interaction

$$B_p = - \quad \begin{matrix} & X & \\ X & & X \\ & X & \end{matrix} \quad . \tag{6.14}$$

This has the desired effect. Placing the interaction at the indicated plaquette performs the following transformation of loops

.

It's not hard to convince yourself that all loop states can be reached from the empty state, so all closed loop patterns must be in the superposition. The final Hamiltonian is

$$H_{\mathrm{TC}} = - \sum_{v \in \mathrm{vertices}} \quad \begin{matrix} & Z & \\ Z & & Z \\ & Z & \end{matrix} \quad - \sum_{p \in \mathrm{plaquettes}} \quad \begin{matrix} & X & \\ X & & X \\ & X & \end{matrix} \tag{6.15}$$

and the ground state is an equal superposition over all closed loop states:

$$\Big| \quad \Big\rangle + \Big| \, \text{\reflectbox{}} \, \Big\rangle + \Big| \, \begin{matrix} \end{matrix} \, \Big\rangle + \Big| \, \Diamond \, \Big\rangle + \Big| \, \Big\rangle + \cdots \tag{6.16}$$

*Note that the Toric code Hamiltonian is usually presented in the $|+\rangle/|-\rangle$ basis rather than the $|0\rangle/|1\rangle$ basis.*

**Toric code ground state**

The simplest way to construct a PEPS for the toric code uses the structure of the ground state. The PEPS tensor is constructed to ensure the superposition of closed loop patterns is achieved upon contraction. The most natural way to achieve this it to write a single tensor for every second plaquette rather than each site.

We begin by adding new edges to the lattice. These edges will become the bonds in the tensor network.

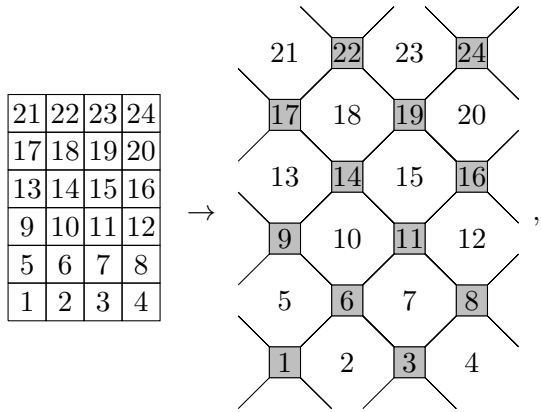

where the plaquettes are numbered for clarity.

Recall that the ground state is built using loops of $|1\rangle$ in a background of $|0\rangle$. We choose the state of the added edges such that the loop pattern is preserved

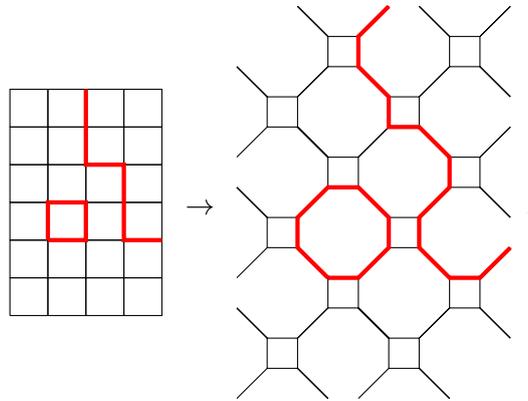

where ▬ indicates a spin in the $|1\rangle$ state on that edge. We choose the following convention when it is ambiguous

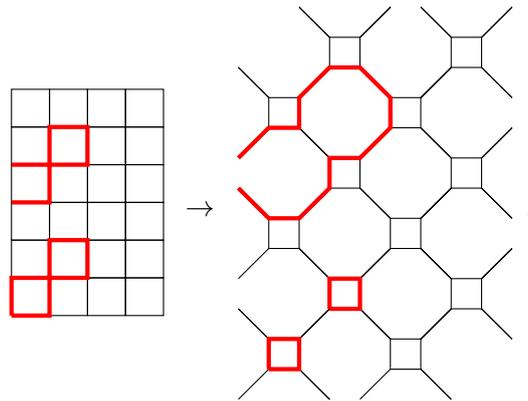

which makes everything consistent.

Interpreting these added edges as bonds in a tensor network, we obtain a PEPS tensor for every second plaquette in the original lattice with four physical indices. The nonzero components are

$$
\begin{array}{c}
l \\
l+i \quad k+l \\
i \qquad\qquad k \quad = 1, \qquad (6.17) \\
i+j \quad j+k \\
j
\end{array}
$$

where $i, j, k, l \in \mathbb{Z}_2$. In this tensor the straight legs indicate virtual indices, and the wavy legs physical indices, specifically the four qubits on the given plaquette. The network looks as below, with the dotted lines representing the original lattice:

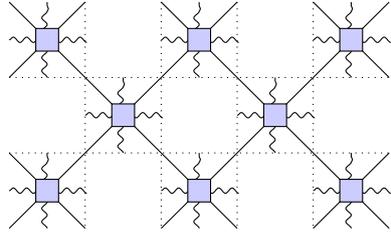

$$(6.18)$$

This tensor simply ensures that if adjacent physical indices are in the $|1\rangle$ state, i.e. carrying a loop, then the virtual index between them does not carry a loop which would leave the plaquette. Conversely, if only one is in the $|1\rangle$ state, the loop must leave the plaquette.

Since an even number of the virtual bonds must be in the $|1\rangle$ state for the tensor entry to be nonzero, the PEPS tensor has a property called $\mathcal{G}$-injectivity [6]. This means that there is a symmetry on the virtual level

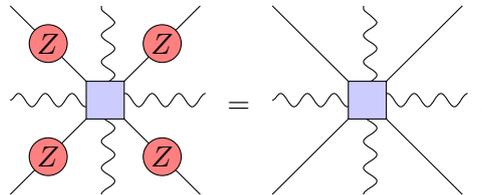

$$(6.19)$$

This turns out to be closely related to the topological order present in this model.

## 6.4 2D Cluster State and the complexity of PEPS

Let $D = 2$ be the bond dimension and let

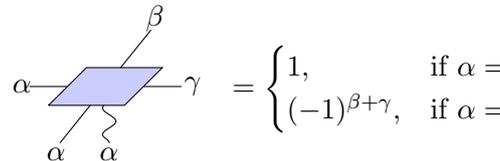

$$(6.20)$$

be the only nonzero elements of the PEPS tensor. The physical state generated is the 2D cluster state, a universal resource for measurement based quantum computing [7,8].

If we could efficiently take the inner product between PEPS (i.e. contract a square grid network), then we can clearly classically simulate single qubit post selected measurements by simply contracting rank 1 projectors onto the physical indices of these PEPS tensors. This shows us that we cannot contract even simple PEPS states efficiently, unless post-selected quantum computing can be classically simulated (Post-BQP=BPP) [9].

### 6.4.1 Numerical PEPS

Although we will not discuss the details of numerical implementation of PEPS algorithms, we note that the status is not as dire as the previous section would imply. In many practical situations, approximate contraction of PEPS networks can be achieved in both the finite [10] and infinite [11,12] system size limits.

## 6.5 Properties of PEPS

Above, we saw a number of properties of 1D PEPS or MPS. We will now see which properties hold in two dimensions. One might naïvely expect MPS and more general PEPS to share similar properties. As we will see below, these two tensor network states share qualitatively different properties, both in terms of the physics the corresponding states exhibit, and in the computational power of the tensor networks.

single tensor

$$i - \boxed{Q} - l \quad = \sum_{s_a} e^{-\frac{\beta}{2}(h[s_i, s_a] + h[s_j, s_a] + h[s_k, s_a] + h[s_l, s_a])}. \tag{6.25}$$

Let

$$H = -J \sum_{\langle i,j \rangle} s_i s_j, \tag{6.26}$$

where $s \in \{\pm 1\}$, the classical Ising model. The tensor $Q$ then simplifies to

$$i - \boxed{Q} - l \quad = 2 \cosh \left( \frac{\beta J}{2} (s_i + s_j + s_k + s_l) \right). \tag{6.27}$$

### 6.5.1 Algebraic decay of correlations

As we saw above, MPS can only capture states with exponential decay of correlations (or constant correlations of course). We will now see if this holds in the case of PEPS. We can build a PEPS state corresponding to a classical partition function by modifying the above construction [3]. Let

$$i - \boxed{D} - l \quad = \delta_{i,j,k,l,x}, \qquad\qquad i - \boxed{M} - j \quad = e^{-\frac{\beta}{2} h[s_i, s_j]}, \tag{6.28}$$

or equivalently combine these into

$$i - \boxed{Q} - l \quad = e^{-\frac{\beta}{4}(h[s_i, s_x] + h[s_j, s_x] + h[s_k, s_x] + h[s_l, s_x])}. \tag{6.29}$$

This defines a PEPS state

$$|\psi\rangle = \qquad\qquad\qquad\qquad\qquad\qquad\qquad\qquad\qquad . \tag{6.30}$$

Note this is a pure state, and *not* a thermal state. It is however not normalised, with $\langle \psi | \psi \rangle = Z$. Correlation functions computed using this state are equal to those computed using classical statistical

physics. Suppose we were to consider a classical model with a thermal phase transition (such as the Ising model above). Such a model will exhibit algebraic decay of correlations at the critical temperature, implying that the corresponding PEPS does as well. Thus we can see that unlike MPS, the states described by PEPS *can* exhibit algebraic decay of correlations.

### 6.5.2 Gauge freedom

The gauge freedom of a PEPS tensor is a simple generalisation of the MPS freedom. As before, we can block tensors together without changing the global state. In addition, we can perform the following transformation (on a translationally invariant PEPS):

$$(6.31)$$

where $N$ and $M$ are invertible matrices.

Recall that in the MPS case, we could use this freedom to bring the tensors into a canonical form. This cannot be done exactly in the case of PEPS, though there do exist numerical methods to bring PEPS into *approximate* canonical forms [13].

---

**Problems 6**

Solutions in accompanying document.

1. What is the PEPS tensor required to build the GHZ state on the honeycomb lattice where spins reside on vertices?

2. Which 2 qubit gate is obtained by contracting the following tensors along the horizontal index?

$$\boxed{u}-k \;\; = \delta_{i,j}\left(\delta_{k,0} + (-1)^i \delta_{k,1}\right), \qquad\qquad z-\boxed{v}\;\; = \delta_{x,y,z}. \qquad (6.32)$$

3. The cluster state can be prepared from the all $|+\rangle$ state by applying $CZ$ between all adjacent spins. Show that Eq. (6.20) indeed gives the cluster state.

   *Hint: Consider the decomposition of a gate given in the above problem.*

4. Investigate how logical operators on the physical spins of the Toric code can be pulled onto the virtual level of the PEPS. Can you see why $\mathcal{G}$-injectivity is so important for topologically ordered PEPS?

5. Convince yourself that evaluating expectation values on the PEPS constructed from a classical partition function indeed reproduces the thermal expectation values.

### 6.7 References

[1] F. Verstraete, J. I. Cirac, and V. Murg, "Matrix Product States, Projected Entangled Pair States, and variational renormalization group methods for quantum spin systems," *Advances in Physics* **57**, 143–224, arXiv:0907.2796, (2009).

[2] R. Orús, "A practical introduction to tensor networks: Matrix product states and projected entangled pair states," *Annals of Physics* **349**, 117–158, arXiv:1306.2164, (2014).

[3] F. Verstraete, M. M. Wolf, D. Perez-Garcia, and J. I. Cirac, "Criticality, the area law, and the computational power of projected entangled pair states.," *Physical Review Letters* **96**, 220601, arXiv:quant-ph/0601075, (2006).

[4] N. Schuch, D. Poilblanc, J. I. Cirac, and D. Pérez-García, "Resonating valence bond states in the PEPS formalism," *Physical Review B* **86**, 115108, arXiv:1203.4816, (2012).

[5] A. Y. Kitaev, "Fault-tolerant quantum computation by anyons," *Annals of Physics* **303**, 2–30, arXiv:quant-ph/9707021, (2003).

[6] N. Schuch, I. Cirac, and D. Pérez-García, "PEPS as ground states: Degeneracy and topology," *Annals of Physics* **325**, 2153–2192, arXiv:1001.3807, (2010).

[7] F. Verstraete and J. I. Cirac, "Valence-bond states for quantum computation," *Physical Review A* **70**, 060302, arXiv:quant-ph/0311130, (2004).

[8] N. Schuch, M. M. Wolf, F. Verstraete, and J. Cirac, "Entropy Scaling and Simulability by Matrix Product States," *Physical Review Letters* **100**, 30504, arXiv:0705.0292, (2008).

[9] N. Schuch, M. Wolf, F. Verstraete, and J. Cirac, "The computational complexity of PEPS," *Physical Review Letters* **98**, 140506, arXiv:quant-ph/0611050, (2008).

[10] M. Lubasch, J. Cirac, and M.-C. Bañuls, "Algorithms for finite projected entangled pair states," *Physical Review B* **90**, 064425, arXiv:1405.3259, (2014).

[11] J. Jordan, R. Orus, G. Vidal, F. Verstraete, and J. Cirac, "Classical simulation of infinite-size quantum lattice systems in two spatial dimensions," *Physical Review Letters* **101**, 250602, arXiv:cond-mat/0703788, (2008).

[12] H. N. Phien, J. A. Bengua, H. D. Tuan, P. Corboz, and R. Orus, "The iPEPS algorithm, improved: fast full update and gauge fixing," *Physical Review B* **92**, 035142, arXiv:1503.05345, (2015).

[13] D. Pérez-García, M. Sanz, C. E. González-Guillén, M. M. Wolf, and J. I. Cirac, "Characterizing symmetries in a projected entangled pair state," *New Journal of Physics* **12**, 025010, arXiv:0908.1674, (2010).

# 7 Multiscale Entanglement Renormalisation Ansatz

MPS are extremely useful for understanding low energy states of 1D quantum models. Despite this, they cannot capture the essential features of some important classes of states. In particular, they cannot reproduce the correlations seen in gapless ground states. Recall that MPS always have exponentially decaying correlations, whereas gapless ground states generically support correlations with power law decay. Similarly MPS also have a strict area law for entanglement entropy, where gapless states admit a logarithmic divergence. The multiscale entanglement renormalisation ansatz is a tensor network designed to overcome these problems.

As mentioned in lecture 5, although MPS do not naturally support the kind of correlations expected in critical models, they have been successfully applied for the study of such systems nonetheless. Using MPS for this purpose requires a family of MPS of increasing bond dimension to examine how the correlations behave. The MERA state functions differently. As we will discuss, a single MERA state can naturally capture the physics of a gapless ground state.

Here, we will present the tensor network as an ansatz and argue that it is well suited to representing ground states of gapless Hamiltonians in 1D. Suppose the state can be written as

$$|\psi\rangle = \quad \text{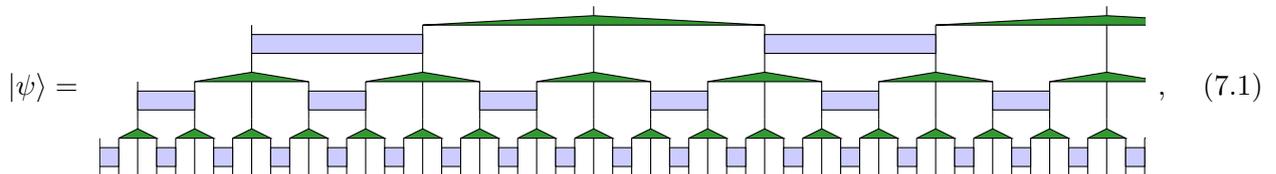} \quad , \quad (7.1)$$

where

$$\text{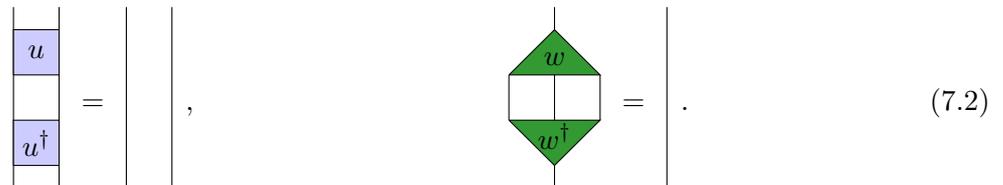} \quad (7.2)$$

As we will see, these constraints on the tensors have both a physical and computational impact. Note that the $u$ and $w$ tensors do not have to be identical, although we frequently restrict to this case if we expect translationally and scale invariant states. The class of states which are expressed as Eqn. 7.1 are known as Multiscale Entanglement Renormalisation Ansatz (MERA) states [1–5].

Although we will not discuss it here, the MERA can be straightforwardly generalised to higher dimensional systems [6–9]. Unlike PEPS, the network can be efficiently optimised in higher dimensions, although the scaling makes the numerics very challenging!

## 7.1 Properties of MERA

### Logarithmic violation of the area law

One of the key properties realised in the MERA which cannot be realised in MPS is a scaling of entanglement entropy. This is easily seen by bond counting. Recall that if $n$ bonds must be broken to separate a region from the rest of the network, the maximum entanglement entropy that can be supported is $n \log D$, where $D$ is the bond dimension. Recall that in the case of MPS any reduced state on a contiguous region can be removed by cutting $n = 2$ bonds.

By inspecting the diagram

$$\text{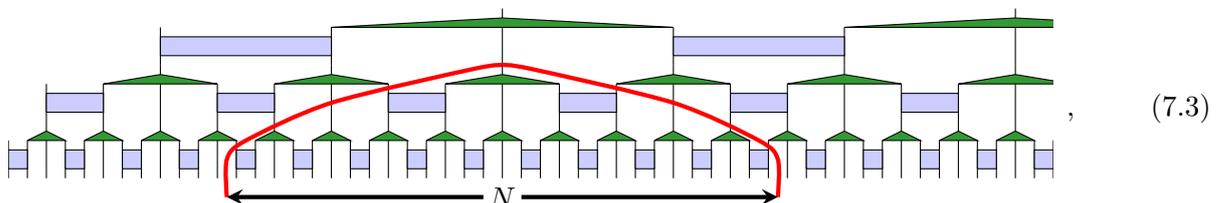} \quad , \quad (7.3)$$

it is straightforward to see that to remove a block of $N$ physical indices from the rest of the network, $\mathcal{O}(\log N)$ bonds must be cut. This shows that the maximum entropy scales as $\log N \log D$ [1, 2].

**Power law decay of correlations**

Using the constraints on the tensors (Eqn. 7.2), we can simplify the evaluation of a two point correlator on a MERA state [3].

$$\langle\psi|O_jO_{j+N}|\psi\rangle = \quad \tag{7.4}$$

$$= \quad \tag{7.5}$$

$$= \quad \tag{7.6}$$

$$= \quad . \tag{7.7}$$

Note that the length scale behaviour of the correlator is completely determined by the application of a superoperator

$$\mathcal{S}(\phi) = \quad , \tag{7.8}$$

where the $w$ tensor can be viewed as a set of Kraus operators

$$M_k = \quad \tag{7.9}$$

obtained by grouping the indices indicated.

Thus, $\mathcal{S}$ is a completely positive, unital map and all eigenvalues $\lambda$ of $\mathcal{S}$ are $|\lambda| \leq 1$. We can bring operators separated by $N$ sites together by applying $\mathcal{S} \sim \log N$ times. Considering eigenoperators of the $\mathcal{S}$ superoperator, the correlator acts as

$$\langle A_j B_k \rangle \sim \frac{\langle A_0 B_1 \rangle}{|j - k|^{\Delta_A + \Delta_B}}, \tag{7.10}$$

where $\Delta_\phi = -\log_3 \lambda_\phi$, $\Delta_\phi \geq 0$ are known as scaling dimensions, where $\lambda_\phi$ is the corresponding eigenvalue of $\mathcal{S}$. Therefore, a MERA state can support algebraic decay of correlations. Although this discussion required the operators to be placed at special sites, it can be easily generalised.

**Efficient Manipulation**

As described in Section 1.5, a good tensor network ansatz should fulfil two properties. First, it should be efficiently storable. All of the networks we have discussed thus far have this property, as only a small number of coefficients are required to represent these states. The second property is more subtle; one should be able to extract physical data efficiently. Although this works for the 1D MPS network, it fails for 2D PEPS states; the contractions required to calculate expectation values of local operators is incredibly hard.

It turns out the MERA has both of these properties. One can efficiently store the state data, and, thanks to the constraints in Eqn. 7.2, one can efficiently compute local expectation values and correlators. We have already seen how this works. The isometric constraints ensure that local operators on the physical level of the network are mapped to local operators on the higher levels [10]. Therefore, computing expectation values only requires manipulation of a small number of tensors in the *causal cone* of the operator

$$\langle O \rangle = \quad \underset{}{\text{}} \quad , \tag{7.11}$$

where the shaded region indicates the causal cone of the five site operator on the physical level indicated in yellow. Notice that the number of tensors on each subsequent level does not grow. Indeed, after a single layer of tensors, the operator becomes a three site operator, and the range never grows. Thus, we see that the layers of the MERA act to map local operators to local operators.

## 7.2 Renormalisation Group Transformation

Much of the discussion above concerned interpretation of the layers of the MERA as Kraus operators, defining a unital CP map on local operators. Evaluating expectation values can be seen as application of many superoperators followed by the inner product with some state on a smaller number of sites

$$\langle \psi_0 | O | \psi_0 \rangle = \langle \psi_{k+1} | \mathcal{A}_k \circ \cdots \circ \mathcal{A}_1 \circ \mathcal{A}_0(O) | \psi_{k+1} \rangle, \tag{7.12}$$

where $\mathcal{A}_j$ is a map from $3^{N-j}$ spins to $3^{N-j}/3$ spins. This can be seen as a renormalisation group or scale transformation. The state $|\psi_j\rangle$ is supported on $3^{N-j}$ spins, and contains only the physical data necessary to understand the physics on that length scale. As we saw, if $O$ is a local operator, $\mathcal{A}(O)$ is easy to evaluate. This allows us to understand the effective operator as a function of length scale [1, 3, 4].

The thermodynamic or macroscopic observables can be seen as the operators obtained by applying a formally infinite number of MERA layers to the high energy or microscopic observables. Thus, the macroscopic physics, or phase structure, is determined by fixed points of the maps $\mathcal{A}$. Some particularly interesting states are the scale invariant states. If the MERA tensors are all the same after some layer, the state is scale invariant. For these states, we do not expect the physics to change as a function of length or energy scale. The fixed point observables of these states are particularly simple to understand, and distinct scale invariant states characterise the different phases.

Since there is no characteristic length scale set by either the spectral gap or correlation length, gapless ground states are expected to be scale invariant. The MERA therefore allows us to understand the long range physics of these states incredibly efficiently [3, 10]. Another way to achieve a scale invariant state is to have zero correlation length — these states characterise gapped phases.

## 7.3 AdS/CFT

In the appropriate limit, the low energy physics of the gapless spin chains considered here is described by a *conformal field theory* (CFT) [12, 13]. The physics of CFTs is thought to be related to gravitational theories in one additional dimension [14–16].

This duality can be observed in the MERA network [17–19]. Imposing the graph metric on the MERA, we find a discretised anti-de Sitter (AdS) metric [17], whilst the edge theory is a 'discretised' CFT. In addition to being a concrete realisation of the holographic principle, the MERA/CFT duality provides avenues towards designing quantum error correcting codes [20].

We note that the AdS/MERA connection remains an open research question. Limits on the ability of MERA states to replicate physics on scales less than the AdS radius have been shown [19]. Additionally, whether the geometry is best understood as anti-de Sitter [17] or de Sitter [18] is currently unclear. Whatever the status, the connection is intriguing. We encourage the interested reader to explore the rapidly expanding literature on the topic [19–28].

## 7.4 Some Simple MERA States

**Product State**

Let

$$
w = \begin{array}{c} \\ 000 \\ 100 \\ 010 \\ 110 \\ 001 \\ 101 \\ 011 \\ 111 \end{array}
\begin{array}{c} \phantom{0} \\ \end{array}
\begin{pmatrix}
1/2 & 0 \\
1/2 & 0 \\
0 & 1/2 \\
0 & 1/2 \\
1/2 & 0 \\
1/2 & 0 \\
0 & 1/2 \\
0 & 1/2
\end{pmatrix}
\begin{array}{c} 0 \quad\quad 1 \end{array}
\tag{7.13}
$$

and $u = \mathbb{1}$.

If we build $\log_3 N$ layers using these tensors, we end up with a state on $N$ sites. The network still has a free index at the top, so we need to define a one-index 'top tensor' $T$ to obtain the final state. Let $T = |+\rangle$. The state obtained is $|+\rangle^{\otimes N}$.

**GHZ State**

Let

$$
\begin{array}{c} l \\ \triangle \\ i \quad j \quad k \end{array} = \delta_{i,j,k,l},
\tag{7.14}
$$

and $u = \mathbb{1}$. Let the top tensor be $T = |+\rangle$. The state obtained is $\frac{|0\rangle^{\otimes N} + |1\rangle^{\otimes N}}{\sqrt{2}}$.

**Cluster State**

It is more convenient to define the cluster state on a binary MERA than a ternary. Place two spins at each site and let

$$
\triangle = \left( \begin{array}{c} \boxed{H} \end{array} \right),
\tag{7.15}
$$

where $\bullet\!\!-\!\!\bullet$ is a controlled-Z gate and H is the Hadamard. If we pick a top tensor $T = |++\rangle$, we obtain the cluster state on periodic boundary conditions.

**Gapless states**

Recently, a family of analytic MERA for the critical point of the transverse field Ising model was proposed [11]. One can also use numerical techniques to obtain a MERA approximation to the ground state of a local Hamiltonian however. Here, we will present some physical data obtained for a model known as the transverse field cluster model [29]. In particular, we will present the ground state energy and the decay exponents ($\Delta_\phi$ in Eqn. 7.10).

This model is most straightforwardly defined with a pair of spin half particles at each site. The Hamiltonian for this model is

$$
\begin{aligned}
H = & -\sum_j \left( X_j^{(1)} + X_j^{(2)} + Z_{j-1}^{(2)} X_j^{(1)} Z_j^{(2)} + Z_j^{(1)} X_j^{(2)} Z_{j+1}^{(1)} \right) \\
& - \lambda \sum_j \left( X_j^{(1)} X_j^{(2)} + Z_j^{(1)} Y_j^{(2)} Y_{j+1}^{(1)} Z_{j+1}^{(2)} \right).
\end{aligned}
\tag{7.16}
$$

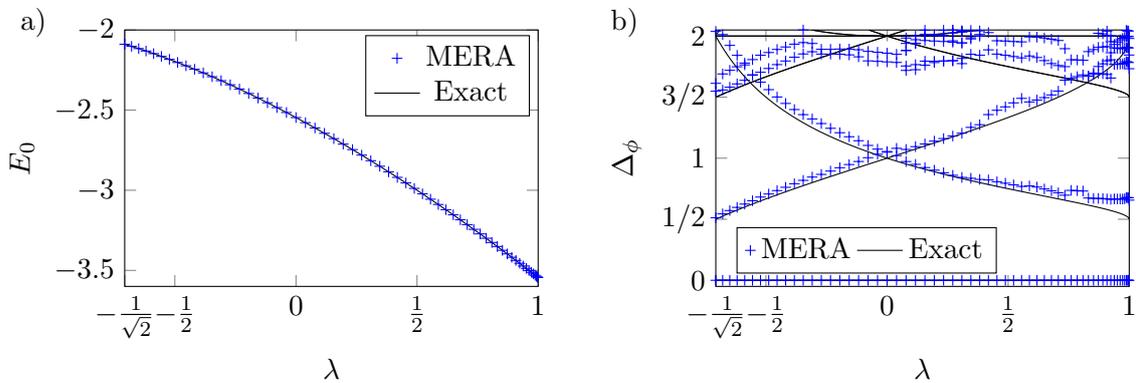

Figure 1: a) Ground state energy density extracted from a ternary MERA after optimising the tensors to locally minimise the energy.
b) Correlation decay exponents for the transverse field cluster model obtained from a ternary MERA.
Figures reproduced from Ref. [29].

This is the cluster state Hamiltonian with transverse fields and an additional interaction with variable strength. The Hamiltonian remains gapless for a range of values of $\lambda$, over which the ground state energy varies continuously as seen in Fig. 1a). The decay exponents also vary over this range, meaning that the thermodynamic physics or RG fixed point is dependent on $\lambda$. These exponents can easily be extracted from an optimised MERA by finding the eigenvalues of the $\mathcal{S}$ superoperator in Eqn. 7.8. The MERA results are shown in Fig. 1b).

---

**Problems 7**

Solutions in accompanying document.

1. Can you find a MERA for the W state?

2. What state is given by the MERA with

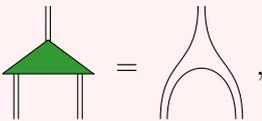

$$(7.17)$$

   $u = \mathbb{1}$ and top tensor $T = \frac{1}{\sqrt{2}}(|00\rangle + |11\rangle)$?

3. The above state is the ground state of the Hamiltonian

$$H = -\sum_{j=1}^{N/2}(X_{2j}X_{2j+1} + Z_{2j}Z_{2j+1}) \qquad (7.18)$$

   on periodic boundary conditions. Is that clear? Can you find a unitary $U_{2j-1,2j}$ which transforms this Hamiltonian into

$$H = -\sum_{j=1}^{N/2}(Z_{2j-1}X_{2j}Z_{2j+1} + Z_{2j}X_{2j+1}Z_{2j+2})? \qquad (7.19)$$

4. Act with the above transformation $U$ on the MERA tensor to obtain another MERA tensor. What is this state?

5. What is the maximum range of thermodynamic observables in a ternary MERA scheme?

6. What does the reduced density matrix on a few sites of the MERA look like? Notice that it corresponds to the top tensor being passed through a CPTP map several times, this is usually called the *descending superoperator*.

7. Do tree tensor networks (i.e. MERA for $u = \mathbb{1}$) have any area law violation on contiguous regions?

---

## 7.6 References

[1] G. Vidal, "Entanglement renormalization," *Physical Review Letters* **99**, 220405, arXiv:cond-mat/0512165, (2007).

[2] G. Vidal, "Class of Quantum Many-Body States That Can Be Efficiently Simulated," *Physical Review Letters* **101**, 110501, arXiv:quant-ph/0610099, (2008).

[3] R. N. C. Pfeifer, G. Evenbly, and G. Vidal, "Entanglement renormalization, scale invariance, and quantum criticality," *Physical Review A* **79**, 040301, arXiv:0810.0580, (2009).

[4] G. Evenbly and G. Vidal, "Algorithms for entanglement renormalization," *Physical Review B* **79**, 149903, arXiv:1201.1144, (2009).

[5] G. Vidal, "Entanglement Renormalization: An Introduction," in *Understanding quantum phase transitions* (L. Carr, ed.), ch. 5, p. 115–138, CRC Press, arXiv:0912.1651, (2011).

[6] G. Evenbly and G. Vidal, "Entanglement renormalization in fermionic systems," *Physical Review B* **81**, 235102, arXiv:0710.0692, (2010).

[7] L. Cincio, J. Dziarmaga and M. M. Rams, "Multi-scale Entanglement Renormalization Ansatz in Two Dimensions: Quantum Ising Model," *Physical Review Letters* **100**, 240603, arXiv:0710.3829, (2008).

[8] M. Aguado and G. Vidal, "Entanglement renormalization and topological order," *Physical Review Letters* **100**, 070404, arXiv:0712.0348, (2008).

[9] G. Evenbly and G. Vidal, "Entanglement renormalization in two spatial dimensions," *Physical Review Letters* **102**, 180406, arXiv:0811.0879, (2009).

[10] G. Evenbly and G. Vidal, "Quantum Criticality with the Multi-scale Entanglement Renormalization Ansatz," in *Strongly Correlated Systems. Numerical Methods* (A. Avella and F. Mancini, ed.), ch. 4, Springer, arXiv:1109.5334, (2013).

[11] G. Evenbly and S. R. White, "Entanglement renormalization and wavelets," *Physical Review Letters* **116**, 140403, arXiv:1602.01166, (2016).

[12] P. Di Francesco, P. Mathieu and D. Sénéchal, Conformal Field Theory. Springer, (1997).

[13] P. Christe and M. Henkel, Introduction to Conformal Invariance and Its Applications to Critical Phenomena. Springer-Verlag, (1993).

[14] J. M. Maldacena, "The Large N Limit of Superconformal Field Theories and Supergravity," *International Journal of Theoretical Physics* **38**, 1113, arXiv:hep-th/9711200, (1998).

[15] E. Witten, "Anti De Sitter Space And Holography," *Advances in Theoretical and Mathematical Physics* **2**, 253, arXiv:hep-th/9802150, (1998).

[16] S. S. Gusber, I. R.. Klebanov and A. M. Polyakov "Gauge theory correlators from non-critical string theory," *Physics Letters B* **428**, 105, arXiv:hep-th/9802109, (1998).

[17] B. Swingle, "Entanglement Renormalization and Holography," *Physical Review D* **86**, 065007, arXiv:0905.1317, (2012).

[18] C. Bény, "Causal structure of the entanglement renormalization ansatz," *New Journal of Physics* **15**, 023020, arXiv:1110.4872, (2013).

[19] N. Bao, C. Cao, S. M. Carroll and A. Chatwin-Davies, "Consistency Conditions for an AdS/MERA Correspondence," *Physical Review D* **91**, 125036, arXiv:1504.06632, (2015).

[20] F. Pastawski, B. Yoshida, D. Harlow and J. Preskill, "Holographic quantum error-correcting codes: Toy models for the bulk/boundary correspondence," *Journal of High Energy Physics* **6**, 149, arXiv:1503.06237, (2015).

[21] G. Evenbly and G. Vidal, "Tensor network states and geometry," *Journal of Statistical Physics* **145**, 891, arXiv:1106.1082, (2011).

[22] M. Nozaki, S. Ryu, and T. Takayanagi, "Holographic Geometry of Entanglement Renormalization in Quantum Field Theories," *Journal of High Energy Physics* **2012**, 193, arXiv:1208.3469, (2012).

[23] B. Swingle, "Constructing holographic spacetimes using entanglement renormalization," arXiv:1209.3304, (2012).

[24] T. Hartman and J. Maldacena, "Time Evolution of Entanglement Entropy from Black Hole Interiors," *Journal of High Energy Physics* **2013**, 014, arXiv:1303.1080, (2013).

[25] M. Miyaji, S. Ryu, T. Takayanagi, and X. Wen, "Boundary States as Holographic Duals of Trivial Spacetimes," *Journal of High Energy Physics* **2015**, 152, arXiv:1412.6226, (2015).

[26] M. Miyaji and T. Takayanagi, "Surface/State Correspondence as a Generalized Holography," *Progress of Theoretical and Experimental Physics* **2015**, 073B03, arXiv:1503.03542, (2015).

[27] M. Miyaji, T. Numasawa, N. Shiba, T. Takayanagi, and K. Watanabe, "cMERA as Surface/State Correspondence in AdS/CFT," *Physical Review Letters* **115**, 171602, arXiv:1506.01353, (2015).

[28] W. -C. Gan, F. -W. Shu, and M. -H. Wu, "Thermal geometry from CFT at finite temperature," *Physics Letters B* **750**, 796, arXiv:1506.01353, (2015).

[29] J. C. Bridgeman, A. O'Brien, S. D. Bartlett, and A. C. Doherty, "Multiscale entanglement renormalization ansatz for spin chains with continuously varying criticality," *Physical Review B* **91**, 165129, arXiv:1501.02817, (2015).

# A PEPOs for local Hamiltonians: The 'particle decay' construction

In numerical algorithms such as DMRG, operators such as Hamiltonians are often represented in the form of Matrix Product Operators (MPO) in 1D, and Projected Entangled Pair Operators (PEPO) in 2D and higher, as seen below. For highly structured Hamiltonians, such as those which are local and translation invariant, an analytic MPO construction of such operators is known in 1D [1]. In this section we review this, and outline a generalisation which allows for local Hamiltonians (and even slightly less structured operators) to be optimally expressed as a PEPOs in arbitrary spatial dimensions.

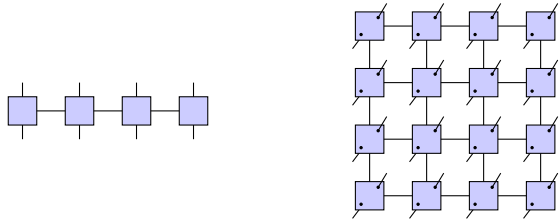

Much like in Eqs. (3.54) and (3.58) we are going to omit the physical indices, as such we will consider MPO tensors to be (operator-valued) matrices, and PEPO tensors to be (operator-valued) rank-$2D$ tensors in $D$ spatial dimensions.

In this section we will need to specify individual tensor values, as well as the values of a tensor network for a specific index designation. For brevity, we will therefore omit the legs in our diagrams, indicating specific entries in a tensor by a ▢ surrounded by the index values. For example the identity is given by $i\,▢\,i = 1$ for all $i$. To make the constructions more clear we will also allow for non-numeric index values, and denote the index set by $I$.

## A.1 1D

In this notation, if we label our indices $I = \{\cdot, 1, \rightarrow\}$, then the transverse Ising model Hamiltonian given in Eq. (3.54) is given by

$$\cdot\,▢\,\cdot\, = \,\rightarrow\,▢\,\rightarrow\, = \mathbb{1} \tag{A.1}$$

$$\rightarrow\,▢\,\cdot\, = -hZ \qquad \rightarrow\,▢\,1 = X \qquad 1\,▢\,\cdot\, = -JX \tag{A.2}$$

where the boundary terms fix the far left and right indices to $|\rightarrow\rangle$ and $|\cdot\rangle$ respectively.

One common interpretation of this construction is in terms of finite-state automata, with the index values corresponding to the automaton states, and the non-zero index values to the transition rules. The automaton moves from left to right[13], with the boundary vectors setting the initial state to $|\rightarrow\rangle$ and final state to $|\cdot\rangle$. With only these restrictions, the automaton can transition from $|\rightarrow\rangle$ to $|\cdot\rangle$ either directly (giving the field term $-hZ$), or via 1 (giving the Ising term $-JXX$) at any location.

To make the higher dimensional generalisation clear we will slightly modify this finite-state automata language, to that of particles and their decay. We can think of $\rightarrow$ as a right-moving particle, and $\cdot$ as the vacuum. The first two transition rules (A.1) correspond to both the vacuum and particle being stable states, with the remaining transitions (A.2) to valid decay routes of the particle. Thus we can interpret the value of the overall MPO as being a superposition over all decays, with each corresponding to a term in the Hamiltonian.

### Heisenberg Model

Suppose we wish to construct a Hamiltonian containing multiple two-body terms, such as the Heisenberg anti-ferromagnet, which contains the terms $-J_X XX$, $-J_Y YY$, $-J_Z ZZ$, as well as a field $-hZ$. An MPO of this model is given in standard notation in Eq. (3.58).

---

[13]Though a right-to-left convention is more commonly used in this 1D construction, a left-to-right convention will prove useful for consistency with the higher dimensional construction.

Added Hamiltonian terms can be accommodated in this construction by extra decay chains. Take our index set to be $I = \{\cdot, x, y, z, \rightarrow\}$ and our MPO to have terms:

$$\cdot\ \square\ \cdot = \mathbb{1} \qquad\qquad \rightarrow\ \square\ \rightarrow\ = \mathbb{1} \qquad\qquad (A.3)$$

$$\rightarrow\ \square\ x = X \qquad\qquad x\ \square\ \cdot = -J_X X \qquad\qquad (A.4)$$

$$\rightarrow\ \square\ y = Y \qquad\qquad y\ \square\ \cdot = -J_Y Y \qquad\qquad (A.5)$$

$$\rightarrow\ \square\ z = Z \qquad\qquad z\ \square\ \cdot = -J_Z Z \qquad\qquad (A.6)$$

$$\rightarrow\ \square\ \cdot = -hZ \qquad\qquad (A.7)$$

Again Equations A.3 correspond to stable vacuum and particles, and each of the transition rules Eqs. (A.4) to (A.7) to each term in the Hamiltonian.

**Cluster Model**

The Cluster Hamiltonian contains three body terms of the form $ZXZ$. Larger terms such as this can be accommodated by longer decay chains. Take an index set $I = \{\cdot, 1, 2, \rightarrow\}$ and include the standard stable vacuum/particle terms as well as

$$\rightarrow\ \square\ 2 = Z \qquad\qquad 2\ \square\ 1 = X \qquad\qquad 1\ \square\ \cdot = Z. \qquad\qquad (A.8)$$

By combining the above two techniques, we can construct arbitrary local Hamiltonians.

## A.2   2D and higher

In higher dimensions we can use a similar construction. Suppose we want to construct a 2D field Hamiltonian, consisting of a $Z$ at every site. Take our index set to be $I = \{\rightarrow, \cdot\}$. Our typical stable vacuum/particle terms that we will always include now become

$$\cdot\ \overset{\cdot}{\underset{\cdot}{\square}}\ \cdot = \rightarrow\ \overset{\cdot}{\underset{\cdot}{\square}}\ \rightarrow\ = \mathbb{1}. \qquad\qquad (A.9)$$

For the field Hamiltonian we need only allow for a simple particle decay of

$$\rightarrow\ \overset{\cdot}{\underset{\cdot}{\square}}\ \cdot\ =\ Z \qquad\qquad (A.10)$$

As for the boundary conditions, along the top, right and bottom boundaries we will once again fix the only non-zero indices to be the vacuum $|\cdot\rangle$. Along the left edge, the boundary condition is a virtual W-state (c.f. Eq. (3.18)) on indices $\{\rightarrow, \cdot\}$, i.e. the equal superposition of all single-particle states. As such we can see that all the non-zero contributions to the Hamiltonian are of the form:

$$
\begin{array}{ccccccccccc}
 & \cdot & & \cdot & & \cdot & & \cdot & & \cdot & \\
\cdot & \square & \cdot & \square & \cdot & \square & \cdot & \square & \cdot & \square & \cdot \\
 & \cdot & & \cdot & & \cdot & & \cdot & & \cdot & \\
\rightarrow & \square & \rightarrow & \square & \rightarrow & \square & \rightarrow & \square & \cdot & \square & \cdot \\
 & \cdot & & \cdot & & \cdot & & \cdot & & \cdot & \\
\cdot & \square & \cdot & \square & \cdot & \square & \cdot & \square & \cdot & \square & \cdot \\
 & \cdot & & \cdot & & \cdot & & \cdot & & \cdot & \\
\end{array}
\ =\ 
\begin{array}{ccccc}
\mathbb{1} & \mathbb{1} & \mathbb{1} & \mathbb{1} & \mathbb{1} \\
\mathbb{1} & \mathbb{1} & \mathbb{1} & Z & \mathbb{1} \\
\mathbb{1} & \mathbb{1} & \mathbb{1} & \mathbb{1} & \mathbb{1} \\
\end{array}
$$

As with 1D, by introducing intermediary states and different decay rules, arbitrary local Hamiltonians in any dimension can be similarly constructed. For example suppose we wanted a 9-body plaquette term of the form:

$$
\begin{array}{ccc}
J & K & L \\
M & N & O \\
P & Q & R \\
\end{array}
$$

Take $I = \{\cdot, 1, 2, \rightarrow\}$ and our non-trivial decay modes to be

$$\rightarrow \begin{smallmatrix} \cdot \\ \square \\ 2 \end{smallmatrix} \, 2 = J, \qquad 2 \begin{smallmatrix} \cdot \\ \square \\ 2 \end{smallmatrix} \, 1 = K, \qquad 1 \begin{smallmatrix} \cdot \\ \square \\ 2 \end{smallmatrix} \, \cdot = L,$$

$$\cdot \begin{smallmatrix} 2 \\ \square \\ 1 \end{smallmatrix} \, 1 = M, \qquad 2 \begin{smallmatrix} 2 \\ \square \\ 1 \end{smallmatrix} \, 1 = N, \qquad 1 \begin{smallmatrix} 2 \\ \square \\ 1 \end{smallmatrix} \, \cdot = O,$$

$$\cdot \begin{smallmatrix} 1 \\ \square \\ \cdot \end{smallmatrix} \, 2 = P, \qquad 2 \begin{smallmatrix} 1 \\ \square \\ \cdot \end{smallmatrix} \, 1 = Q, \qquad 1 \begin{smallmatrix} 1 \\ \square \\ \cdot \end{smallmatrix} \, \cdot = R.$$

then we can see that the non-zero contributions to the Hamiltonian are of the form

$$
\begin{array}{ccccccc}
\mathbb{1} & \mathbb{1} & \mathbb{1} & \mathbb{1} & \mathbb{1} & \mathbb{1} & \mathbb{1} \\
\mathbb{1} & \mathbb{1} & J & K & L & \mathbb{1} & \mathbb{1} \\
= \quad \mathbb{1} & \mathbb{1} & M & N & O & \mathbb{1} & \mathbb{1} \\
\mathbb{1} & \mathbb{1} & P & Q & R & \mathbb{1} & \mathbb{1} \\
\mathbb{1} & \mathbb{1} & \mathbb{1} & \mathbb{1} & \mathbb{1} & \mathbb{1} & \mathbb{1}
\end{array}
$$

## A.3 Other examples

Below are several more example of Hamiltonian constructed by the above method.

### Toric code (Wen Plaquette)

$$I = \{\cdot, 1, \rightarrow\} \qquad \rightarrow \begin{smallmatrix} \cdot \\ \square \\ 1 \end{smallmatrix} \, 1 = 1 \begin{smallmatrix} 1 \\ \square \\ \cdot \end{smallmatrix} \, \cdot = X \qquad 1 \begin{smallmatrix} \cdot \\ \square \\ 1 \end{smallmatrix} \, \cdot = \cdot \begin{smallmatrix} 1 \\ \square \\ \cdot \end{smallmatrix} \, 1 = Y$$

### Quantum Compass model/Bacon-Shor code

$$I = \{\cdot, 1, \rightarrow\} \qquad \rightarrow \begin{smallmatrix} \cdot \\ \square \\ \cdot \end{smallmatrix} \, 1 = 1 \begin{smallmatrix} \cdot \\ \square \\ \cdot \end{smallmatrix} \, \cdot = X \qquad \rightarrow \begin{smallmatrix} \cdot \\ \square \\ 1 \end{smallmatrix} \, \cdot = \cdot \begin{smallmatrix} 1 \\ \square \\ \cdot \end{smallmatrix} \, \cdot = Y$$

### 2D Transverse Ising

$$I = \{\cdot, 1, \rightarrow\} \quad \rightarrow \begin{smallmatrix} \cdot \\ \square \\ \cdot \end{smallmatrix} \, \cdot = hZ \quad \rightarrow \begin{smallmatrix} \cdot \\ \square \\ \cdot \end{smallmatrix} \, 1 = \rightarrow \begin{smallmatrix} \cdot \\ \square \\ 1 \end{smallmatrix} \, \cdot = JX \quad 1 \begin{smallmatrix} \cdot \\ \square \\ \cdot \end{smallmatrix} \, \cdot = \cdot \begin{smallmatrix} 1 \\ \square \\ \cdot \end{smallmatrix} \, \cdot = -X$$

### 2D Cluster state

$$I = \{\cdot, 1, 2, \rightarrow\} \qquad \rightarrow \begin{smallmatrix} \cdot \\ \square \\ \cdot \end{smallmatrix} \, 2 = 1 \begin{smallmatrix} \cdot \\ \square \\ \cdot \end{smallmatrix} \, \cdot = \cdot \begin{smallmatrix} 1 \\ \square \\ 1 \end{smallmatrix} \, \cdot = \cdot \begin{smallmatrix} \cdot \\ \square \\ 1 \end{smallmatrix} \, \cdot = Z \qquad 2 \begin{smallmatrix} 1 \\ \square \\ 1 \end{smallmatrix} \, 1 = X$$

## A.4 References

[1] U. Schollwöck, "The density-matrix renormalization group in the age of matrix product states," *Annals of Physics* **326**, 96–192, arXiv:1008.3477, (2011).